\documentclass[usenatbib]{mn2e}
\usepackage{psfig}

\def\gs{\mathrel{\raise0.35ex\hbox{$\scriptstyle >$}\kern-0.6em
\lower0.40ex\hbox{{$\scriptstyle \sim$}}}}
\def\ls{\mathrel{\raise0.35ex\hbox{$\scriptstyle <$}\kern-0.6em
\lower0.40ex\hbox{{$\scriptstyle \sim$}}}}

\def\kms{\,\hbox{km}\,\hbox{s}^{-1}}
\def\Msol{\mathrel{\rm M_{\odot}}}

\def\Msolyr{\mathrel{\rm M_{\odot}yr^{-1}}}
\def\Wm2{\,\hbox{W}\,\hbox{m}^{-2}}
\def\gsim{\mathrel{\raise0.35ex\hbox{$\scriptstyle >$}\kern-0.6em
\lower0.40ex\hbox{{$\scriptstyle \sim$}}}}
\def\lsim{\mathrel{\raise0.35ex\hbox{$\scriptstyle <$}\kern-0.6em
\lower0.40ex\hbox{{$\scriptstyle \sim$}}}}

\def\sfir{S_{850}}
\def\srad{S_{1.4}}

\begin{document}

\title[The Properties of Sub-mm Galaxies in Hierarchical Models]{The
  Properties of Sub-mm Galaxies in Hierarchical Models}

\author[Swinbank et al.]
{ \parbox[h]{\textwidth}{ 
A.\,M.\ Swinbank$^{\, 1,*}$,
C.\,G.\ Lacey$^{\, 1}$,
Ian Smail$^{\, 1}$,
C.\,M.\ Baugh$^{\, 1}$,
C.\,S.\ Frenk$^{\, 1}$,\\
A.\,W.\ Blain$^{\, 2}$,
S.\,C.\ Chapman$^{\, 3}$,
K.\,E.\,K.\ Coppin$^{\, 1}$,
R.\,J.\ Ivison$^{\, 4,5}$,
L.\,J.\ Hainline$^{\, 6}$,
J.\,E.\ Gonzalez$^{\, 1}$
}
\vspace*{6pt} \\
$^1$Institute for Computational Cosmology, Department of Physics, Durham University, 
South Road, Durham, DH1 3LE, UK \\
$^2$Astronomy Department, California Institute of Technology, 105-24, Pasadena, CA 91125, USA \\
$^3$Institute of Astronomy, University of Cambridge, Madingley Road, Cambridge, CB3 0HA, UK \\
$^4$UK Astronomy Technology Center, Royal Observatory, Blackford Hill, Edinburgh, EH19 3HJ, UK \\
$^5$Institute for Astronomy, University of Edinburgh, Edinburgh, EH19 3HJ, UK \\
$^6$Department of Astronomy, University of Maryland, College Park, MD 20742, USA \\
$^*$Email: a.m.swinbank@durham.ac.uk \\
}
\maketitle 

\begin{abstract}
  We use the combined {\sc galform} semi-analytical model of galaxy
  formation and {\sc grasil} spectrophotometric code to investigate the
  properties of galaxies selected via their sub-millimeter (sub-mm)
  emission.  The fiducial model we use has previously been shown to fit
  the properties of local ULIRGs, as well as the number counts of faint
  sub-mm galaxies.  Here we test the model in more detail by comparing
  the SEDs and stellar, dynamical, gas and halo masses of sub-mm
  galaxies against observational data.  We precisely mimic the sub-mm
  and radio selection function of the observations and show that the
  predicted far-infrared properties of model galaxies with
  $S_{850}>5$\,mJy and $S_{1.4}>30\mu$Jy are in good agreement with
  observations.  Although the dust emission model does not assume a
  single dust temperature, the far-infrared SEDs are well described by
  single component modified black-body spectrum with characteristic
  temperature $32\pm5$\,K, in good agreement with observations.  We
  also find evidence that the observations may have uncovered evolution
  in the far-infrared--radio relation in ULIRGs out to $z\sim 2$.  We
  show that the predicted redshift distribution of sub-mm galaxies
  provides a reasonable fit to the observational data with a median
  redshift $z=2.0$.  The radio-selected subset of sub-mm galaxies are
  predicted to make up approximately 75\% of the population and peak at
  $z=1.7$, in reasonable agreement with the observed radio detected
  fraction and redshift distribution.  However, the predicted $K$-band
  and mid-infrared (3--$8\mu$m) flux densities of the sub-mm galaxies
  (and LBGs) are up to a factor $10\times$ fainter than observed.  We
  show that including the stellar TP-AGB phase in the stellar
  population models does not make up for this deficit. This discrepancy
  may indicate that the stellar masses of the sub-mm galaxies in the
  model are too low: M$_{\star}\sim10^{10}$\,M$_{\odot}$, while
  observations suggest more massive systems,
  M$_{\star}\gsim10^{11}$\,M$_{\odot}$.  However, if the predicted $K$-
  and 3--8$\mu$m extinctions in the model could be dramatically
  reduced, then this would reduce, but not eliminate, this discrepancy.
  Finally we discuss the potential modifications to the models which
  may improve the fit to the observational data, as well as the new
  observational tests which will be made possible with the arrival of
  new facilities, such as SCUBA2.
\end{abstract}


\section{Introduction}

The discovery a decade ago of a population of faint submm-selected
galaxies (SMGs) revolutionised our view of the cosmic star formation
history of the Universe \citep{Smail97,Hughes98,Barger98}.  These
galaxies (originally discovered with the SCUBA instrument at 850$\mu$m)
appear to be high redshift galaxies with star-formation rates exceeding
1000$\Msolyr$ \citep{Ivison00,Smail02}.  These Ultra-Luminous Infrared
Galaxies (ULIRGs) peak around $z\sim2$ \citep{Chapman03a,Chapman05a}
and show a thousand fold increase in their abundance between $z=0$ and
$z=2$.  If this far-infrared emission arises solely from star-formation
with a standard (``solar neighborhood'') IMF, then they could potentially
dominate the star-formation activity in the early Universe, dwarfing
the contribution of galaxies selected in the rest-frame ultraviolet.
The apparent intensity of these starbursts, the resulting high
metallicity, along with their large dynamical masses, high gas
fractions and inferred strong clustering
\citep{Blain04a,Greve05,Tacconi06,Tacconi08,Swinbank04,Swinbank06b} are
all suggestive of a close link to the formation phase of the most
massive spheroids and black holes
\citep{Lilly99,Smail02,Smail04,Webb03,Genzel03,Smail04,Alexander03a,Alexander05a,Swinbank06b}.

Reproducing the sub-mm galaxy population has been a major challenge
for theoretical models \citep{Granato00,Baugh05}.  In part this is
because (quite reasonably) many of the recipes and constraints used to
develop the models are based on the formation and evolution of
``normal'' galaxies rather than the extreme populations.  Hence these
models have had difficulty reproducing extremely luminous galaxies with
sufficient cool dust at high redshift without over-predicting the
abundance of bright galaxies in the local Universe.  Initial attempts
to match the basic properties of sub-mm galaxies (whilst maintaining
the match to the present day $K$-band luminosity function and {\it
  IRAS} 60$\mu$m luminosity function) under-predicted the 850$\mu$m counts by a
factor $30\times$ \citep{Baugh05} despite $\Lambda$CDM producing
enough baryons in massive halos at $z\sim2$ to match observations of
gas masses in sub-mm galaxies \citep{Genzel03,Greve05}.

One solution to this problem was to alter the Initial Mass Function
(IMF): \citet{Blain99b} suggested that a Salpeter IMF
\citep{Salpeter55} with a low-mass threshold of 3\,M$_{\odot}$ was
required in far-infrared luminous galaxies around $z\sim2$ since the
implied star-formation rate for a Salpeter IMF integrated to the
canonical value of 0.07\,M$_{\odot}$ would over-predict the integrated
stellar density at $z=0$.  Invoking a top-heavy (or ``flat'') IMF in
the semi-analytic framework has been shown to provide a much improved
fit to the sub-mm galaxy counts and redshift distribution as well as
the Lyman-break galaxy luminosity function whilst still maintaining a
good fit to the properties of the present day galaxy populations
(\citealt{Baugh05}, hereafter B05).  In this model, a standard IMF is
adopted in quiescently star-forming galaxies, whilst the
star-formation induced by galaxy mergers produces stars with a flat
IMF: $dn/dln(m)=m^{-x}$ with $x=0$ (compared to a Salpeter
IMF which has $x=1.35$; \citealt{Salpeter55}).  With a larger
proportion of high mass stars, the energy radiated in the ultra-violet
per unit mass of stars produced is increased, thus increasing the
amount of radiation to heat the dust.  Moreover, the flat IMF produces
a higher yield of metals from type II supernovae, thus increasing the
dust content of the galaxy and boosting the luminosity in the
sub-millimeter waveband.  

This assumption of a flat IMF in bursts is controversial and so it is
essential to explore the predictions of this model in more detail.  In
particular, the model can be (i) tested against the growing
observational multi-wavelength data on sub-mm galaxies to test whether
the choice of parameters is suitable and (ii) provide useful
constraints on observational data, in particular for understanding
possible selection biases.  For example, since much of what is known
about sub-mm galaxies is based on the radio-detected sub-sample, one
outstanding issue is how much this radio-selection (which does not
benefit from the negative K-correction experienced in the sub-mm
waveband) has affected the conclusions being drawn about SMGs.  Are
the radio-undetected fraction of SMGs at either significantly higher
redshift \citep{Younger07} or do they instead have far-infrared
colours which represent much colder systems \citep{Chapman05a}?

In \S2 we describe the basic properties of the galaxy formation model
we employ, but refer the reader to \citet{Cole2k},
\citet{Benson02,Benson03}, \citet{Baugh05} and \citet{Lacey08} for a
detailed description.  In \S3 we test the predicted far-infrared
properties of sub-mm-selected galaxies against the available
observational data.  In \S4 we discuss the masses and evolution of
model SMGs compared to those estimated from observational data and in \S5 we give
our conclusions and future prospects for constraining the properties of
sub-mm galaxies both theoretically and observationally.  We use a flat,
$\Lambda$CDM cosmology with $\Omega_{m}=0.3$, $\sigma_8=0.93$ and
$H_{o}=100h$\,km\,s$^{-1}$\,Mpc$^{-1}$ with $h=0.7$.

\section{The  Model}

The galaxy formation model which we employ is the one described in
detail in \citet{Baugh05} and \citet{Lacey08}.  The model is based on
the semi-analytic code of \citet{Cole2k} with important revisions
described in \citet{Benson02,Benson03} and is reviewed extensively in
\citet{Baugh06}.  In summary, the formation, assembly and evolution of
galaxies are calculated using a background cosmology in which  the
dark matter structure grows hierarchically.  The physical ingredients
considered in the model include: (i) The formation of dark matter halos
through mergers and accretion of material. (ii) The collapse of baryons
into the gravitational potential wells of dark matter halos.  (iii) The
radiative cooling of gas that is shock heated during infall into the
dark halo. (iv) The formation of a rotationally supported disk of cold
gas. (v) The formation of stars from the cold gas. (vi) The injection
of energy into the interstellar medium, through supernova explosions or
the accretion of material onto a super-massive black hole.  (vii) The
chemical evolution of the interstellar medium, stars and the hot gas.
(viii) The merger of galaxies following the merger of their host dark
matter halos, due to dynamical friction.  (ix) The formation of
spheroids during mergers due to the rearrangement of pre-existing stars
(i.e. the disk and bulge of the progenitor galaxies) and the formation
of stars in a burst. (x) The construction of a star-formation history
and a composite stellar population for each galaxy.

The spectral energy distribution (SED) from this composite stellar
population is calculated using the spectrophotometric model {\sc
  grasil} \citep{Silva98}.  {\sc grasil} computes the emission from
both the stars and dust in a galaxy, based on the star formation, metal
enrichment history and the sizes as predicted by the semi-analytical
model \citep{Granato00}.  {\sc grasil} includes radiative transfer
through a two-phase dust medium, with a diffuse component and giant
molecular clouds, and a distribution of dust grain sizes.  We note that
{\sc grasil} does not assume a single dust temperature for the galaxy.
Stars are assumed to form inside the clouds and then gradually migrate
out.
The output from {\sc grasil} is the galaxy SED from the far-UV to radio
wavelengths.  

\subsection{Radio Emission and the Far-Infrared--Radio Correlation}
\label{sec:farir-radio_corr}

In addition to the sub-mm/far-infrared emission from dust, the radio
emission is also included in the model following \citet{Bressan02}.
The radio emission is produced by (i) thermal bremsstrahlung from H{\sc
  ii} regions and (ii) synchrotron radiation powered by acceleration of
relativistic electrons in supernova remnants. In local galaxies, there
is a strong correlation between non-thermal radio and far-infrared
emission, holding over five decades of luminosity (e.g.\
\citealt{Helou85,Yun01,Vlahakis07}).  The standard explanation of this
relationship is that both the far-infrared and the bulk of the radio
emission are caused by high-mass ($\gsim5$\,M$_{\odot}$) stars.  These
stars both heat the dust (which then emits far-infrared emission) and
at the end of their lives explode as supernovae (see e.g.
\citealt{Condon92} and references therein), producing the relativistic
electrons responsible for synchrotron radiation.  The strength of this
synchrotron emission depends on not only the number density of
relativistic electrons, but also the strength of the magnetic field.
The rate of energy input into the relativistic electrons depends on the
assumed low-mass cut off for Type~II supernova production, which is
usually taken as 8\,M$_{\odot}$ \citep{Bressan02}.

In the model galaxies, both the thermal and non-thermal contributions
to the radio emission are calculated.  The thermal component of the
radio emission is estimated from the stellar emission essentially
without free parameters, assuming that all ionizing photons are
absorbed by gas within the galaxy, and is proportional to the
instantaneous ionizing luminosity of the stars.  The synchrotron
(non-thermal) radio emission is assumed to be proportional to the
instantaneous Type~II supernova rate, but its calculation involves two
empirical parameters, which are essentially the efficiency with which
the supernova explosion energy is converted into energy of relativistic
electrons, and the power-law index of the injection energy spectrum of
these relativistic electrons, which determines the spectral index of
the synchrotron emission. It is assumed that the radiative lifetime of
the relativistic electrons is short enough that the total synchrotron
luminosity is always equal to the instantaneous energy injection rate
from supernovae. \citet{Bressan02} estimate both of these parameters
from observations of local galaxies; in particular, the efficiency
factor was chosen based on observations of the Milky Way galaxy.  They
found that this choice also reproduced the observed ratio of radio to
far-infrared emission for normal spirals. In this model, the ratios
between the far-infrared and radio luminosities and the instantaneous
star formation rate (SFR) all approach constant values if the SFR
varies only on timescales $\gsim$100\,Myr (as in normal
spirals). However, due to the different lifetimes of the stars powering
the far-infrared, thermal and non-thermal radio emission, there are
timelags between the corresponding luminosities and the formation of
the stars involved, and so these ratios vary if the SFR changes on
timescales $\lsim$100Myr, as is the case in starbursts. This is
discussed in detail in \citet{Bressan02}, and has consequences for the
results in this paper.

\begin{figure}
\centerline{\psfig{file=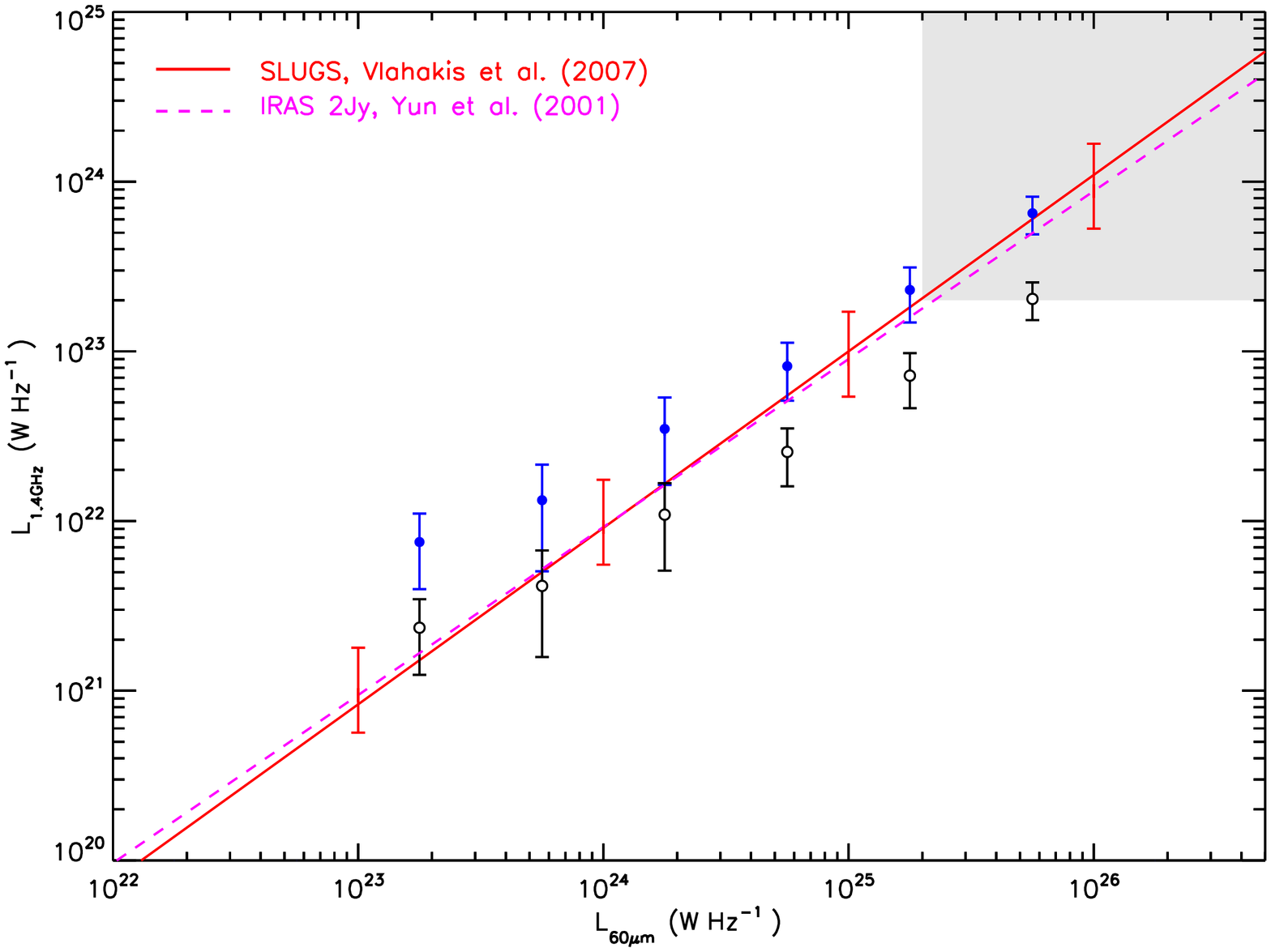,angle=90,width=3.5in}}
\caption{Far-Infrared -- radio correlation for galaxies in the {\sc
    grasil} model compared to the SCUBA Local Galaxy Survey (SLUGS) and
  IRAS 2\,Jy survey from \citet{Yun01} and \citet{Vlahakis07}.  The red
  error-bars show the scatter in the observed data.  The shaded area
  denotes the approximate 60$\mu$m and radio luminosities encompassing
  the ULIRGs from \citet{sanders03}.  The model ($z<0.1$) predictions
  are overlaid for both the original radio luminosities (open circles)
  and after the renormalisation factor $3.2\times$ has been applied to
  the radio flux densities (filled circles).  We adopt this
  renormalisation of the model to ensure that it reproduces the
  far-infrared and radio properties of local galaxies with luminosities
  comparable to those observed for high-redshift SMGs -- giving us a
  reliable local benchmark.}
\label{fig:FIR-radio}
\end{figure}

In Fig.~\ref{fig:FIR-radio} we show the far-infrared--radio (60$\mu$m
versus 1.4\,GHz) correlation for local star-forming galaxies and
overlay the model predictions for galaxies at $z<0.1$.  This shows that
the model reproduces the form of the observed far-infrared--radio
correlation at lower luminosities.  However, at higher luminosities the
predicted radio--far-infrared relation lies below that observed
\citep{Helou85,Yun01,Vlahakis07}, by a factor $3.2\pm0.2\times$,
assuming the \citet{Bressan02} normalisation of the radio emission
using the Milky Way.  As we are interested in the evolution of SMGs,
which have ULIRG-like luminosities (L$_{bol}>10^{12}$L$_{\odot}$) and
appear likewise to be dusty starbursts, it seems appropriate to
renormalize the non-thermal radio emission in the model so that it
matches the properties of high-luminosity {\it IRAS}-selected samples
at $z=0$. This seems likely to produce the most realistic estimates for
the radio luminosities of SMGs, in the framework of the present galaxy
formation model. Physically, this renormalization corresponds to
assuming a higher efficiency for conversion of supernova blastwave
energy into energy of relativistic electrons in starbursts as compared
to normal spirals, which might result from the higher densities or
magnetic fields in starbursts. In the present paper, we have therefore
multiplied the non-thermal radio luminosities from {\sc grasil} by a
factor 3.2, as shown in Fig.~\ref{fig:FIR-radio}.  We note that this
renormalization might not be needed at $z=0$ in a galaxy formation
model which predicts significantly different burst timescales in
ULIRGs.  Indeed, \citet{Bressan02} reproduced the SED of the ULIRG
Arp220, including its radio emission, using their standard synchrotron
normalization, but treating the burst timescale as a free parameter.
However, in our galaxy formation model, the burst timescales are
already fixed by other considerations.

In what follows, we parameterize the relative strength of the
far-infrared and radio emission using $q_L$, defined as
$q_L=log(L_{bol}/([4.52\,{\rm THz}]\,L_{1.4\,\rm GHz}))$ where
$L_{bol}$ is the total dust luminosity and $L_{1.4\,\rm GHz}$ is the
rest-frame radio luminosity. The latter is calculated from the observed
1.4GHz flux $S_{1.4}$ using $L_{1.4\,\rm GHz}=4\pi
D_L^2S_{1.4}(1+z)^{\alpha-1}$, where the spectral index, $\alpha$, has
a typical value of 0.7 for non-thermal sources; \citealt{Condon92}).
For local ULIRGs, $q_L$ is measured to be $q_L=2.34\pm0.01$ with
$\sigma_{q_L}=0.10$ (e.g.\ \citealt{Yun01}).  As described above, we
have normalised the radio emission of $z<0.1$ model ULIRGs so that by
construction they have $q_L=2.34$. The model predicts a dispersion
$\sigma_{q_L}=0.11$.  We stress that this renormalisation ensures that
our model agrees with observations of the radio and far-infrared
emission of galaxies in the luminosity range appropriate for SMGs in
the local Universe.  However, the model also predicts evolution in
these quantities.  For an exponentially declining star-formation rate
with a top-heavy IMF, the radio flux density will plateau shortly after
the initial burst since the SNe are produced from stars with mass
$\gsim8$\,M$_{\odot}$ which have lifetimes of approximately 50\,Myrs.
In contrast, the far-infrared luminosity will continue to increase for
longer as dust is produced and then heated by new generations of stars
extending down to lower masses $\gsim5$\,M$_{\odot}$. As a result, the
value of $q_L$ is not constant for bursts, but depends both on the SFR
timescale and on the age at which the burst is observed
\citep{Bressan02}.

The average $q_{L}$ in bursts therefore depends both on redshift and
on how the galaxy sample is selected.  For model ULIRGs at $z=2$
(which overlap substantially with SMGs), we find $q_{L}=2.19\pm0.05$
and $\sigma_{q_{L}}=0.26$ .  Moreover, the e-folding timescale for
star formation in model ULIRGs more than halves between $z=0$ and
$z=2$ ($\tau_{eff}=300_{-150}^{+250}$\,Myr at $z=0$ compared to
$130_{-60}^{+150}$\,Myr at $z=2$).  In addition, the median age of the
burst in model ULIRGs at $z=2$ is much shorter than at $z=0$
($8_{-6}^{+50}$ compared to $50_{-40}^{+400}$\,Myr).  The young ages
of the high-redshift model ULIRGs, combined with the time-lag between
the onset of the burst and the production of the first supernova, thus
results in an  evolution in the far-infrared--radio relation
for these luminous galaxies.  We will return to this point when
studying the radio and far-infrared properties of SMGs later.

\section{Analysis}

\subsection{Simulating Sub-mm Catalogues}

To test the {\sc galform/grasil} model we need to compare its
predictions to the detailed multi-wavelength properties of
high-redshift sub-mm galaxies.  To achieve this we have to focus on
samples with reliable galaxy identification (which has so far relied on
high resolution radio imaging), and follow-up spectroscopy to provide
precise and unambiguous redshifts.  By far the largest and most secure
redshift survey of sub-mm galaxies comes from (\citealt{Chapman05a}
hereafter C05) who surveyed a total area of approximately 0.25
degree$^2$ across seven fields, securing spectroscopic redshifts for 73
sub-mm galaxies with 850$\mu$m flux densities $\geq5\mu$Jy and radio
counterparts with flux densities $S_{1.4}\geq30\mu$Jy.  This comprises
the primary comparison sample for our analysis.

The C05 survey relies on radio identification as the only secure
method to pin-point the counterparts of large samples of the sub-mm
emitting galaxies.  This is because the large beam-size ($\sim15''$
for the JCMT at 850$\mu$m) conspires with the large number of possible
counterparts making it impossible to identify the galaxy responsible
for the far-infrared emission with 850$\mu$m imaging alone;
\citep{Smail00,Ivison98,Ivison00,Ivison02,Ivison05,Ivison07}.  Using
the deepest radio imaging currently available
($\sigma\sim5$--$10\mu$Jy), this radio selection identifies
$\sim65$--80\% of the bright ($\sfir >5$\,mJy) sub-mm galaxies
\citep{Ivison05,Ivison07}, but potentially introduces biases when
extrapolating the properties of radio identified sub-mm galaxies to
the whole population.  We investigate the effect of this bias in the
model by defining two classes of sub-mm galaxy: first, all sub-mm
galaxies with $\sfir >5$\,mJy are called SMGs; second, we define the
subset of this population which are detectable in the radio (with
$\sfir >5$\,mJy and $\srad >30\mu$Jy) as radio-identified sub-mm
galaxies or rSMGs.

In order to simulate observational sub-mm catalogues, we must model
the measurement noise.  In particular, since most sub-mm catalogues
are cut at a threshold around 3.5--$4\sigma$, sub-mm maps suffer from
{\it flux boosting} in which the flux limit of low significance
sources can be increased above the survey signal-to-noise due to (a)
confusion (the contribution of fainter sources within the large beams)
and (b) the inclusion of low-significance sources which are boosted
above the survey signal-to-noise due to the coincidence alignment with
positive noise spikes.  In order to properly compare the observations
with the models we therefore convolve the model sub-mm and radio flux
densities with a typical 1-$\sigma$ noise of $\sigma_{850}=1.5$\,mJy
(e.g.\ \citealt{Scott02,Coppin07}) and $\sigma_{1.4}=5$--$10\mu$Jy
respectively (e.g.\ \citealt{Ivison02,Ivison07}), thus allowing a
like-for-like comparison with observations.  This measurement noise is
assumed in all following sections and analysis, and is included in all
of the estimates which we give for average masses and luminosities of
model galaxies.

\subsection{Number Counts, Redshift Distribution and Space Density of SMGs}

\begin{figure}
\centerline{\psfig{file=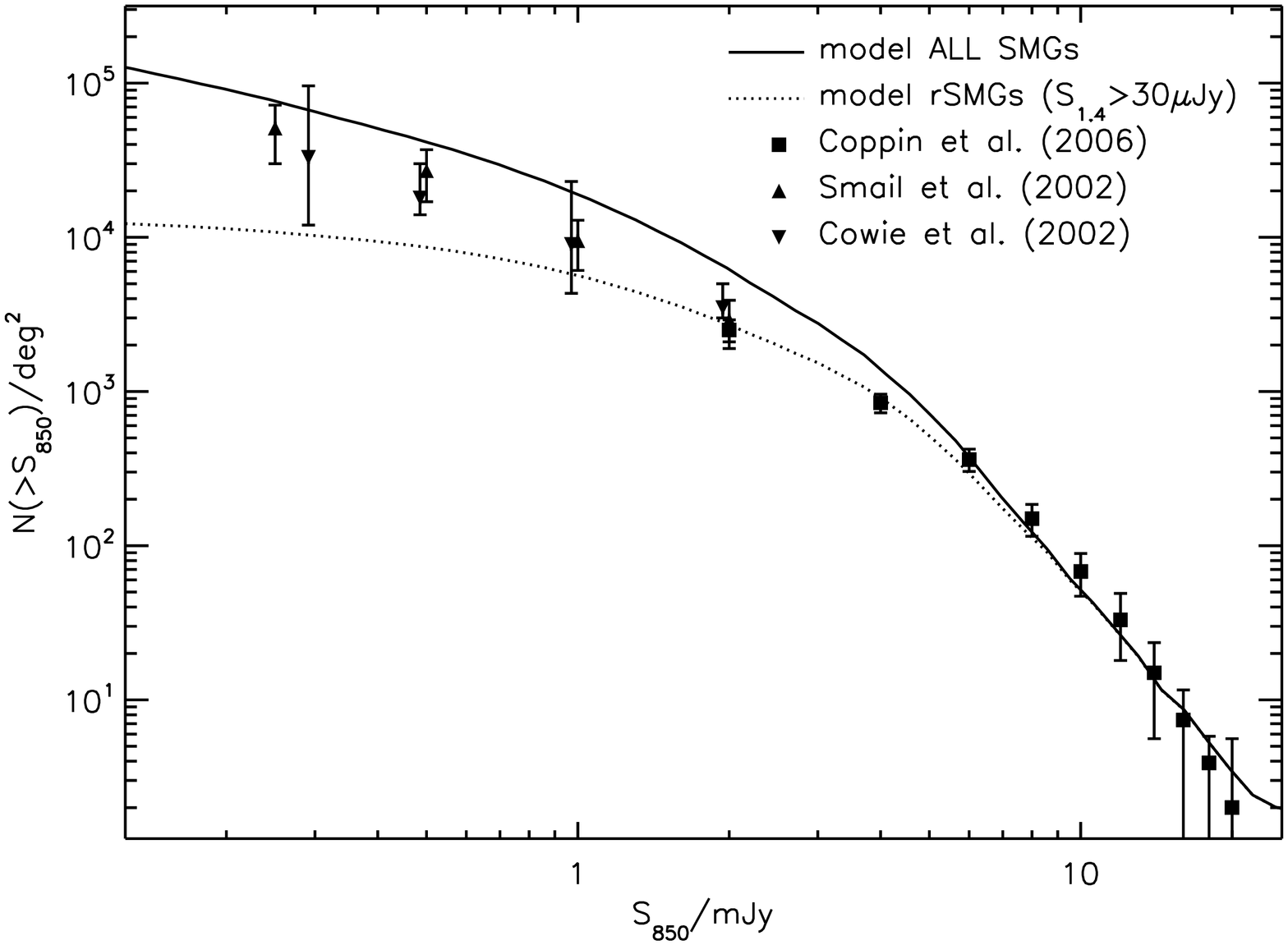,angle=90,width=3.5in}}
\caption{The cumulative number counts per degree$^{2}$ at 850$\mu$m as
  a function of sub-mm flux density.  The model predictions are shown
  for SMGs and rSMGs respectively.  We also plot the observational
  results from various surveys with SCUBA on the JCMT
  \citep{Smail02,Cowie02,Coppin06}.  Above 5\,mJy, the model suggests
  that the surface density of sub-mm galaxies with radio flux densities
  above 30$\mu$Jy is 75\%.  This agrees well with the observed
  radio-detected fraction of S$_{850}>$5\,mJy, S$_{1.4}>$30$\mu$Jy
  sub-mm galaxies of 65-80\% \citep{Ivison05,Ivison07}.  }
\label{fig:counts_radiocuts}
\end{figure}

In Fig.~\ref{fig:counts_radiocuts} we show the predicted cumulative
source counts as a function of 850$\mu$m flux density compared to the
observations derived from a number of different surveys with SCUBA.
The fiducial model of B05 agrees well with the observations down to the
deepest flux density limits, with the model counts above 0.1\,mJy being
dominated by ongoing merger-driven bursts at high-redshift, as these
authors demonstrated.

To determine the effect of the radio-preselection on the number counts,
we also show the cumulative source counts for galaxies which also have
radio flux densities $S_{1.4}>30\mu$Jy.  At a characteristic 850$\mu$m
flux density limit of 5\,mJy of C05, the model suggests that the radio
detected fraction is $\sim75$\%, consistent with the fraction typically
found in sub-mm surveys (65--80\%) \citep{Ivison05,Ivison07}.

The redshift distribution of the radio-identified SMGs from C05
(crudely corrected for spectroscopic incompleteness due to the
redshift desert) is well fit by a Gaussian profile with a median
redshift $z=2.0$ and $\sigma_z=0.7$.  Using a simple model to account
for radio incompleteness, C05 estimate the underlying redshift
distribution of SMGs to peak at $z=2.2$ with a 1-$\sigma$ width of
$\sigma_z=1.3$.

\begin{figure}
\centerline{\psfig{file=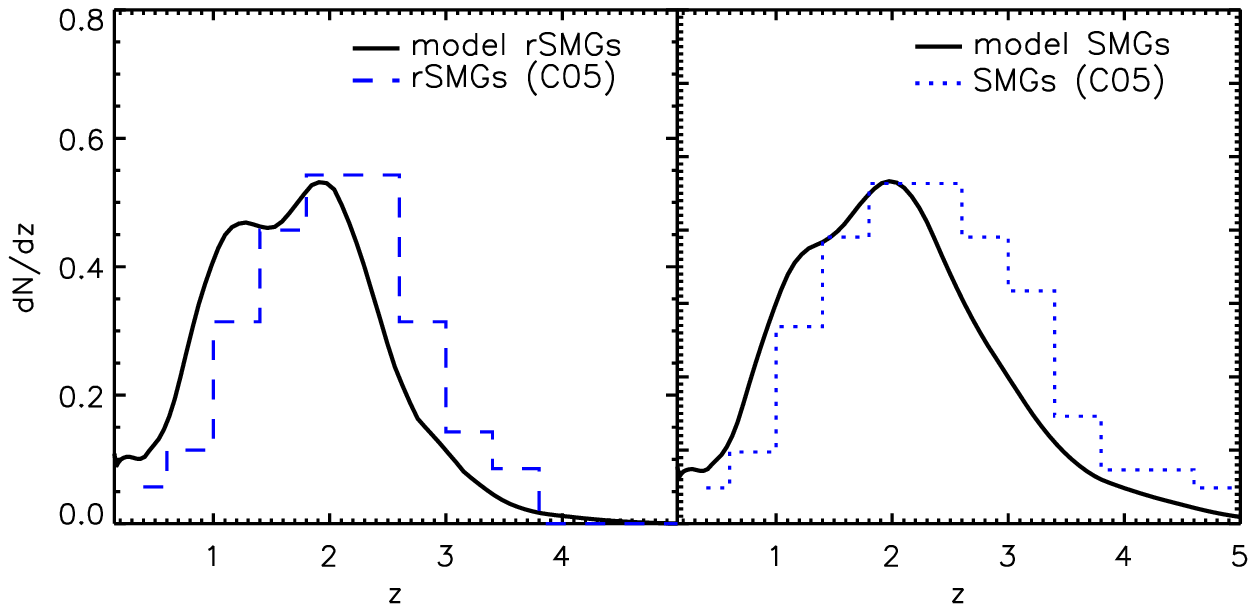,angle=90,width=3.2in}}
\caption{The predicted redshift distribution of model SMGs and rSMGs.
  {\it Left:} $N(z)$ for galaxies selected at both sub-mm and radio-
  wavelengths. The median redshift in the model rSMGs is $z=1.7$ with
  $\sigma_z=0.8$. We over-plot the redshift distribution of rSMGs from
  \citet{Chapman05a} (corrected for spectroscopic incompleteness in the
  redshift desert). {\it Right:} $N(z)$ for a pure sub-mm selected
  sample of galaxies with 850$\mu$m flux densities greater than 5\,mJy,
  peaking at $z=2.0$ with $\sigma_z=1.0$. Again we over-plot the
  redshift distribution from \citet{Chapman05a}, corrected for
  spectroscopic incompleteness and radio incompleteness using a simple
  model for the evolution of the radio luminosity function. In both
  panels, the redshift distribution for the model galaxies is slightly
  shallower than the observations suggest (by $\Delta z\sim0.2$--0.3),
  although since the field-to-field variance in the observations is
  $\Delta z\sim0.25$ we conclude that (within the observational
  uncertainties) the redshift distributions are in good agreement.}
\label{fig:Nz}
\end{figure}

In Fig.~\ref{fig:Nz} we show the redshift distribution of SMGs and
rSMGs in the fiducial model of B05 compared to observations.  The
model SMGs have a median redshift of $<\! z\! >=2.0$ with
$\sigma_z=1.0$.  However, as Fig.~\ref{fig:Nz} also shows, the
radio-identified sub-set peaks at $<\! z\! >=1.7$ with $\sigma_z=0.8$.
Although there is reasonable overlap between the model and
observations, the model SMGs and rSMGs both appear to peak at slightly
lower redshift than inferred observationally ($\Delta z\sim0.2$--0.3
in both cases).  
We note that the field-to-field variation between the seven sub-fields
in the C05 sample is $\Delta z\sim0.25$.  This may be a simple
reflection of the fact that SMGs are highly clustered \citep{Blain04a},
and so we expect that the redshift distribution from C05 is likely to
be uncertain by at least this amount due to cosmic variance. Therefore
we conclude that the model and observations are in good agreement.

It is also worth noting that \citet{Clements08} derive a median
redshift $z\sim1.5$ using photometric redshifts (rising to $z\sim1.9$
when including the far-infrared photometry) for a similar sample of
radio and mid-infrared identified SMGs from the SHADES submm survey.
However, as \citet{Clements08} note, there are large errors on
photometric redshifts for sub-mm galaxies ($\Delta z\sim0.5$--1),
especially for the sources with the faintest counterparts.
Nevertheless, their results are also consistent with the model redshift
distributions for the rSMGs.

\begin{figure}
\centerline{\psfig{file=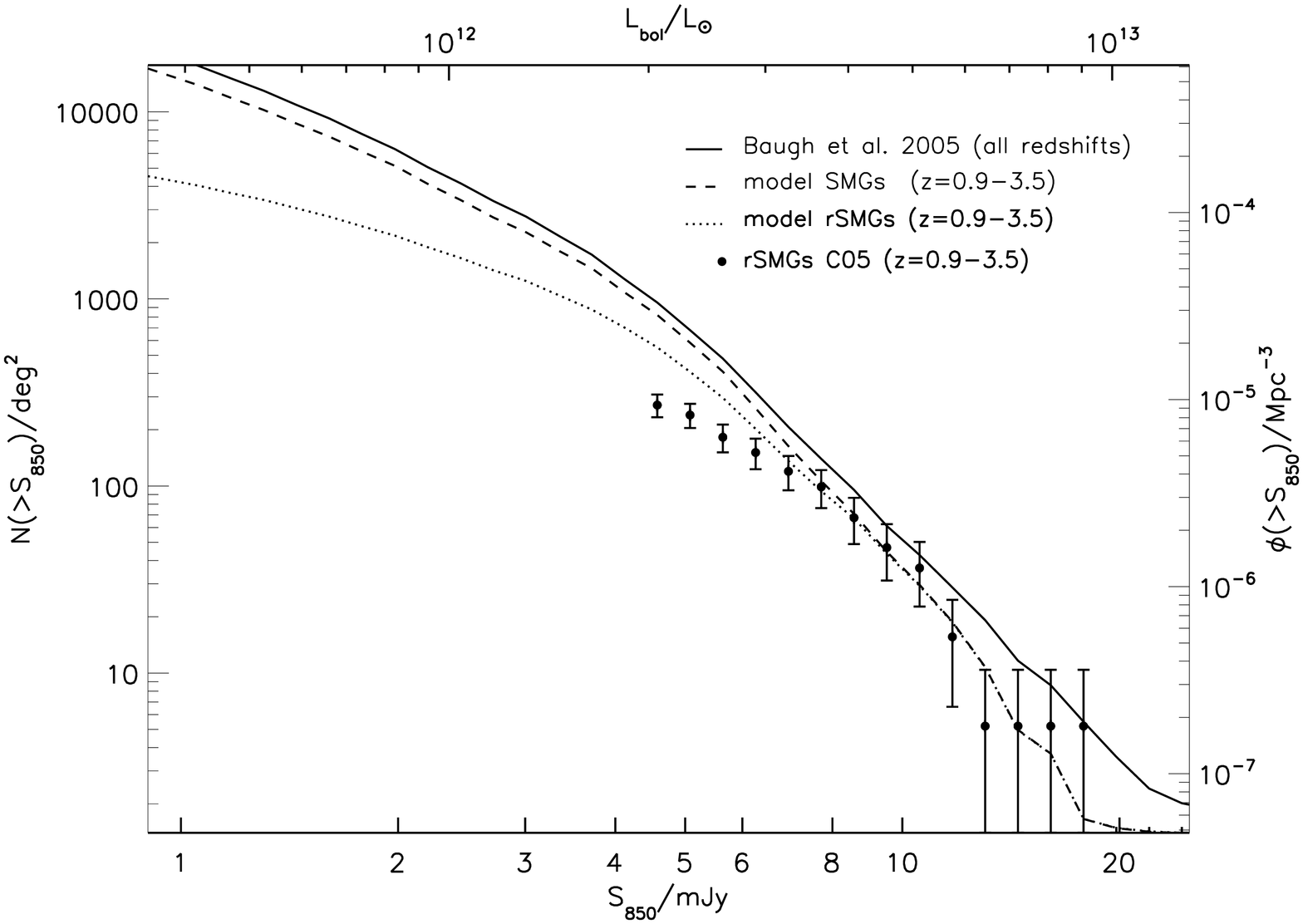,angle=90,width=3.5in}}
\caption{The cumulative number counts per degree$^{2}$ at 850$\mu$m
  for galaxies between $z=0.9$--3.5.  The model predictions including
  all redshifts are shown by the solid line.  We also show the expected
  counts for radio-flux density cuts of $S_{1.4}$\,=0 and 30$\mu$Jy,
  respectively, only considering galaxies in the range $z=0.9$--3.5.
  These are compared to the observational counts for SMGs from the C05
  redshift survey which are radio selected with
  S$_{1.4}\gsim\,30\mu$Jy, and also in the range $z=0.9$--3.5.  The
  right-hand axis denotes the approximate space density in physical
  units, whilst the top axis denotes the approximate bolometric
  luminosity for galaxies at $z=2.0$ given their sub-mm flux
  density.}
\label{fig:counts_rSMGs}
\end{figure}

Using the number counts and redshift distribution, we can also compare
the space densities.  The observed space density of rSMGs from C05
indicates that between $z=0.9$--3.5 the volume density should be
$\sim8.0\times10^{-6}$\,Mpc$^{-3}$.  In comparison, the predicted space
density for model rSMGs in the same redshift interval is
$1.1\pm0.1\times10^{-5}$\,Mpc$^{-3}$, slightly higher but consistent
with the observations.
In Fig.~\ref{fig:counts_rSMGs} we show the space densities over the
redshift interval $z=0.9$--3.5 for the model SMGs compared to the SMGs
in the C05 sample.  To interpret this data as a cumulative luminosity
function, we also convert the number counts and 850$\mu$m flux densities
to approximate space densities and bolometric luminosities assuming a
median redshift $z=2.0$. 

\subsection{Far-infrared Spectral Properties}

Constraining the far-infrared SED of sub-mm galaxies is a key
observational goal since the thermal emission from cold dust dominates
the bolometric luminosity of these galaxies (L$_{8-1000\mu
  m}\gsim0.95$\,L$_{bol}$). Sub-mm photometry (at wavelengths other
than 850$\mu$m) can be used to constrain the SED, measuring the
apparent temperature of the dust, and infer their 
far-infrared luminosity and thus star-formation rate (assuming an
IMF). These diagnostics also allow these galaxies to be placed in the
context of other populations of high-redshift star-forming galaxies
and active galactic nuclei (AGN).

To test how well the SEDs predicted by the model galaxies reproduce the
dust emission spectra of SMGs, we compare the predicted far-infrared
colours with existing observational data.

\subsubsection{1200/850$\mu$m Colours} First, we compare the
850$\mu$m flux density ratio with longer wavelength photometry.  In
particular, it has been claimed that the millimeter/sub-millimeter
flux density ratio is sensitive to redshift above $z=3$
\citep{Eales03,Greve04,Ivison05}.  We therefore compare the
far-infrared photometry predicted by the model SMGs and rSMGs with
recent observational constraints from \citet{Greve08} who surveyed the
GOODS-North region at 1200$\mu$m with the Max-Planck bolometer array
(MAMBO). For an 850$\mu$m selected sample, the model galaxies have a
median 850/1200$\mu$m flux density ratio of
$S_{850}/S_{1200}=2.4\pm0.2$ between $z=1.5$--3, and
$S_{850}/S_{1200}=2.2\pm0.2$ for a 1200$\mu$m selected sample with
$\sigma_{1200}=0.7$\,mJy over the same redshift range (we note that
the error-bars denote the 17 and 84\%-ile range, which corresponds to
the 1-$\sigma$ scatter for a Gaussian distribution). Both the 850- and
1200$\mu$m-selected samples show very little evolution in colour over
the redshift range $z=0.5$--5 and are in good agreement with
observational constraints (Fig.~\ref{fig:fircolours}).

\subsubsection{350/850$\mu$m Colours} A better test of the far-infrared
properties comes from shorter wavelengths (at $z=2$ a black-body with a
characteristic temperature of T$_{d}$=35\,K peaks at an observed
wavelength of $\sim 300\mu$m).  Prior to the launch of {\it Herschel},
the most promising route to constrain shorter wavelengths in the
far-infrared are 350- and 450$\mu$m photometry where the atmospheric
transmission allows the brightest SMGs to be detected in good
conditions.

In Fig.~\ref{fig:fircolours} we compare the predicted
S$_{350}$/S$_{850}$ colours with observational constraints.  At
$z\sim2$ the median flux density ratio for model SMGs is
$S_{350}/S_{850}=2.5\pm2.0$.  For comparison, \citet{Kovacs06} and
\citet{Coppin08} study a total of 27 SMGs at 350$\mu$m which have
secure redshifts.  These SMGs have a median flux density ratio
$S_{350}/S_{850}=4.0^{+4.5}_{-2.2}$ (this value includes upper limits
for 350$\mu$m flux densities on three non-detections).  Since the data
at 350 and 450$\mu$m suffer from both low number statistics and low
signal-to-noise detections we bin the data into two redshift bins,
$z<2$ and $z\geq2$.  As this figure shows, the model predictions seem
broadly consistent with the observational data. There is a hint
($<1.5\sigma$) that the model photometry and data have opposite trends
with redshift, but firm conclusions cannot be drawn with the current
observational data.  However, we stress that this potentially powerful
test will be significantly improved with observations of larger
samples at higher signal-to-noise from upcoming surveys with SCUBA2
and {\it Herschel}.  In the meantime we note that the model 
predicts an evolution in the S$_{350}$/S$_{850}$ flux density ratio
for the SMGs and rSMGs is well described by
S$_{350}$/S$_{850}=-0.6+11.6\times(1+z)^{-1.15}$.

A more useful comparison of the observational and model SEDs can be
made by comparing the bolometric luminosities estimated from fits to
both real and model photometry. To do this, we fit the 350- and
850$\mu$m photometry with a modified blackbody of the form
$L_{\nu}\propto B_{\nu}(T_{d})\nu^{\beta}$ where $B_{\nu}$ is the
Planck function evaluated at the emitted frequency $\nu$, and
$\beta=1.5$ \citep{Dunne03,Coppin08}. The result of this fit is a
characteristic dust emission temperature $T_{d}$ for each galaxy, and
also an estimate of its bolometric dust luminosity $L_{bol}$.
\citet{Kovacs06} (see also more recently \citealt{Coppin08})
demonstrate that sub-mm galaxies have characteristic temperatures
consistent with local starbursts ($T\sim35$\,K), but are at least an
order of magnitude more luminous in $L_{bol}$. We can apply the same
fitting procedure to the 350- and 850$\mu$m photometry for all our
model SMGs. In this way, we find a median characteristic emission
temperature for both model SMGs and rSMGs of $T_{d}=32\pm5$\,K. (Note
that simply a characteristic temperature, since {\sc grasil} does not
assume a single dust temperature even within a single galaxy.) We also
find that the single modified black-body fit (with $\beta=1.5$) yields
an estimated bolometric luminosity which on average is 0.94$\pm$0.25 of
the true model value (integrated between rest-frame 8 and 1000$\mu$m).
We find that the median bolometric luminosity for model SMGs is
$2.0\pm1.5\times10^{12}$\,M$_{\odot}$. Hence it appears that the
far-infrared SEDs of model SMGs are broadly similar to what is inferred
from the observations.

With the bolometric luminosities of the model SMGs in hand, we can
also investigate how the star-formation rates compare to those
inferred observationally. The main difference in the star-formation
rates comes from the adoption of the flat IMF in bursts. This results
in a much lower star-formation rate per unit bolometric luminosity for
SMGs (which are dominated by merger induced bursts).  We find that the
relation between bolometric luminosity and instantaneous
star-formation rate is
SFR(0.15--125\,M$_{\odot}$)(M$_{\odot}$\,yr$^{-1}) =
1.01\times10^{-44}$\,L$_{bol}$\,(erg\,s$^{-1}$)) which means that an
average model SMG with L$_{bol}=2\times10^{12}$\,L$_{\odot}$ has a
star-formation rate of SFR=77\,M$_{\odot}$\,yr$^{-1}$ compared to a
typically observationally-derived value of
SFR(0.1--100\,M$_{\odot}$)=\,$4.5\times10^{-44}$\,L$_{8-1000\mu
m}$\,(erg\,s$^{-1}$) using the \citet{Kennicutt98} calibration for a
Salpeter IMF.

\subsection{Radio properties}
Although the 1.4\,GHz radio emission is used to identify the galaxy
responsible for the sub-mm emission, the sub-mm/radio flux density
ratio has also been used as a diagnostic of both temperature and
redshift \citep{CarilliYun99}.  However, since the far-infrared colours
scale with $(1+z)/T_d$, redshifting a fixed SED template has the same
effect as changing the temperature at a fixed redshift, and so without
knowledge of the temperature, far-infrared colours cannot be
unambiguously used to derive redshifts.  Indeed, with a secure redshift
for a sub-mm galaxy, the effects of radio identification become
apparent: a canonical 5\,mJy radio-identified SMG at $z=2.4$ with a
50$\mu$Jy radio counterpart has a characteristic dust temperature of
32\,K. An increase or decrease in the dust temperature of just 10\,K
has a dramatic (factor $10\times$) effect on the sub-mm and radio-flux
densities and therefore target selection: as the temperature increases
for a fixed luminosity, the hotter SEDs mean that the 850$\mu$m flux
density falls below the detection threshold of 5\,mJy.  Similarly, the
coolest SMGs at $z=2.4$ would have lower radio-flux densities making an
SMG undetectable in the radio \citep{Chapman04c,Blain04a,Blain04c}.

\begin{figure}
\centerline{\psfig{file=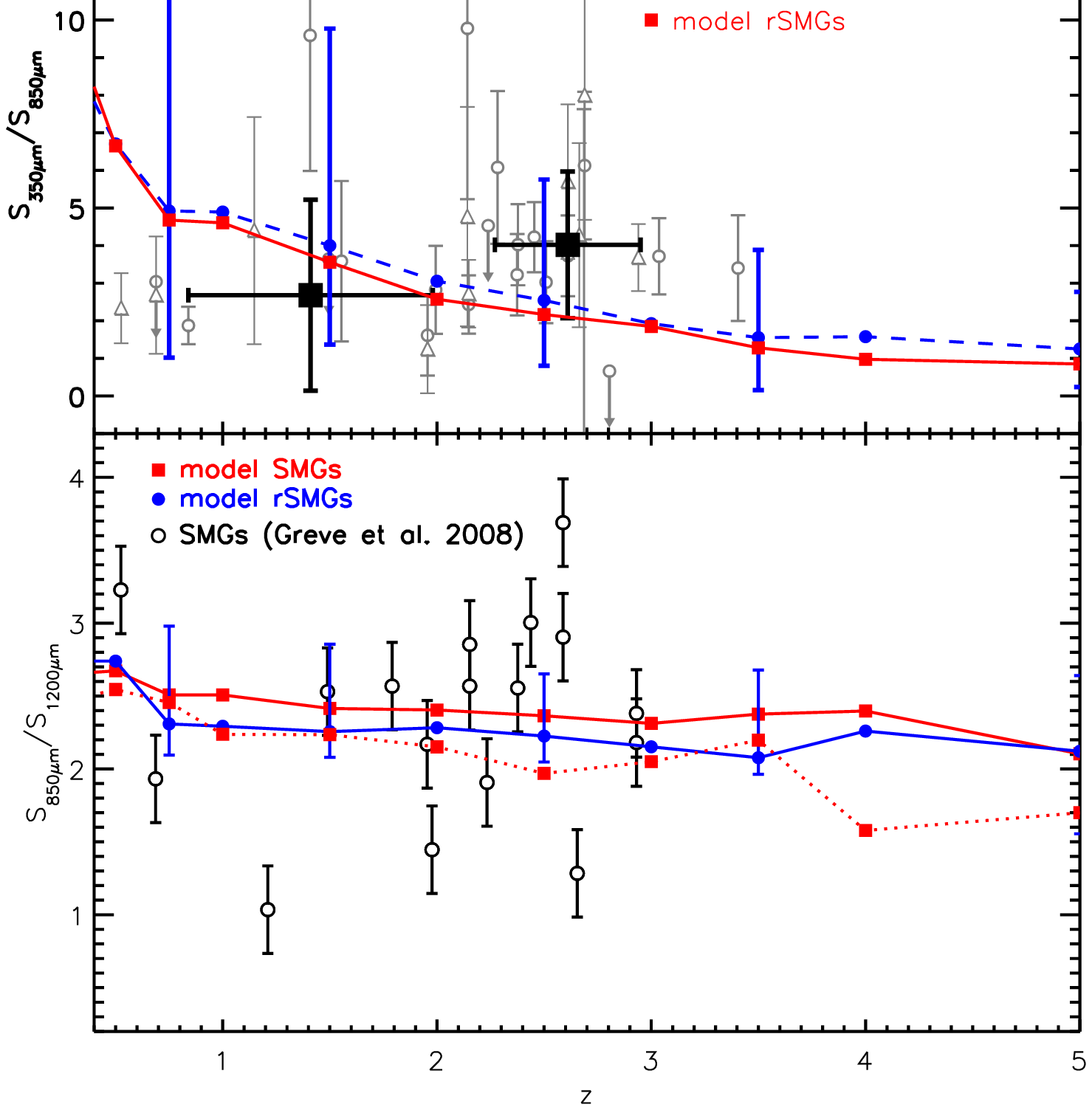,angle=90,width=3.5in}}
\caption{Predicted sub-mm and radio flux density ratios for sub-mm
  galaxies as a function of redshift.  {\it Top:} S$_{850}$/S$_{1.4}$
  colours of sub-mm galaxies in the C05 sample compared to model SMGs
  and rSMGs.  We plot individual galaxies in the C05 sample and show
  the median values in redshift intervals of $\Delta z=0.5$.  We also
  plot the predictions for model SMGs and rSMGs.  The error-bars
  indicate the the typical 67\%-ile of the model predictions, and, for
  clarity are shown on alternate points.  The model rSMGs have a
  systematically larger ($\sim1.26\pm0.24\times$) S$_{850}$/S$_{1.4}$
  flux density ratio than observations suggest over the redshift range
  $z=1$--4. {\it Middle:} S$_{350}$/S$_{850}$ colours as a function of
  redshift for model SMGs and rSMGs.  The comparison sample is taken
  from observations at 350$\mu$m from \citet{Kovacs06} and
  \citet{Coppin08} which have good ($>3\sigma$) detections at
  350$\mu$m.  Within the observational errors the model SMGs are in
  good agreement with observations (which suffer from both low number
  statistics and low-signal to noise detections). To show the redshift
  trend in the observational data more clearly, we bin the data into
  two redshift bins ($z<2$ and $z>2$) (solid squares). {\it Bottom:}
  S$_{850}$/S$_{1200}$ colours as a function of redshift for model SMGs
  and rSMGs compared to observations of the GOODS-North regions by
  \citet{Greve08}.  The solid lines denote the evolution for an
  850$\mu$m selected sample, whilst the dotted line represents the
  evolution for a 2.5\,mJy selected sample at 1200$\mu$m with
  $\sigma_{1200}=0.7$\,mJy.  Both the model SMGs, rSMGs and 1200$\mu$m
  selected model SMGs show remarkably little evolution with redshift
  from $z=0.5$--5.  }
\label{fig:fircolours}
\end{figure}

In Fig.~\ref{fig:fircolours} we compare the predicted $\sfir/\srad$
flux density ratio.  Both the observations and model galaxies show
strong evolution with redshift (a result of strong K-corrections and
evolution).  The model SMGs have a median flux density ratio
$\sfir/\srad=150^{+130}_{-85}$ at $z\sim2$--3, and so a model SMG with
S$_{850}=5$\,mJy has $\srad=30\mu$Jy.  In the model, the
S$_{850}$/S$_{1.4}$ for model SMGs evolves as
S$_{850}$/S$_{1.4}=-200+196\times(1+z)^{0.53}$.  

Whilst the model SMGs appear to reproduce the general evolutionary
trend shown by the observations, over the redshift range $z=0.5$--3.5
they also have a $\sfir/\srad$ flux density ratio which is
systematically too high compared to the observations,
$1.26\pm0.24\times$ (where the error is a bootstrap estimate), despite
matching the $z<0.1$ far-infrared--radio correlation for ULIRGs.  For
the model rSMGs, we derive $<\,q_L\,>=2.19\pm0.08$ with
$\sigma_{q_{L}}=0.20$ -- and we find that the radio and submm flux
selection do not influence this value.  In terms of observations,
\citet{Kovacs06} examine the far-infrared--radio correlation for
fifteen high redshift SMGs at $z\sim2$ and derive
$<\,q_{L}\,>=2.14\pm0.07$ with an intrinsic spread $\sigma_q\sim0.12$.
Compared to ultraluminous far-infrared galaxies at $z=0$, which have
$q_L=2.34^{+0.11}_{-0.10}$ (see \S\ref{sec:farir-radio_corr}), we see
that the model predicts a similar level of evolution to that seen in
the observations.  More importantly, the fact that $q_L$ does not
change for the model SMGs when we include the observational biases
suggests that the low value of $q_L$ found by \citet{Kovacs06} is not a
result of sample selection.  Instead, it appears that the
\citet{Kovacs06} result reflects {\it real} evolution in the
far-infrared--radio correlation.  Whether this results from the same
starburst age-related evolution found in the model, or from other
processes, such as modest contributions to the observed radio flux
densities from AGN activity, requires further observational work.

\subsection{Restframe UV Properties}

With precise positions and redshifts available for the
radio-identified SMGs in C05, this sample has also been an important
basis for more detailed study of the morphologies, stellar populations
and stellar/dynamical and halo mass
\citep[e.g.][]{Smail04,Chapman05a,Swinbank04,Swinbank05b,Swinbank06b,Borys05,Greve05,Pope06,Tacconi06}.
These studies rely on a combination of high resolution imaging with
{\it HST}, as well as multi-band photometry and spectroscopy from
optical, near- and mid-infrared wavelengths.  In particular, the
stellar populations have been probed using optical, near- and
mid-infrared colours to constrain the stellar SED from the rest-frame
UV to near-infrared.

The next step in our comparison is therefore to examine the stellar
populations and masses of model SMGs. We start by comparing the
predicted $B$- and $R$-band magnitudes of SMGs to those from C05.  At
the median redshift of the C05 sample, the observed $B$- and $R$-bands
sample the rest-frame UV, which is very sensitive to the instantaneous
SFR, as well as the level of dust extinction. For model SMGs in the
redshift range $z=0.9$--3.5 the mean $B$- and $R$-band magnitudes are
$B_{AB}=24.0\pm1.3$ and $R_{AB}=23.9^{+2.0}_{-1.0}$. In comparison the
SMGs from C05 have $B_{AB}=24.8\pm1.2$, $R_{AB}=24.3\pm2.0$, slightly
fainter and redder than the models predict but with significant overlap
(just considering the median $(B-R)$ colours, we note that an
additional obscuration of $E(B-V)=0.2$ in the model galaxies would make
the model $B-R$ colours agree with observations).  The model SMGs have
significant overlap in their observed optical colours with BX/BM
galaxies: 45\% of model SMGs have colours consistent with BX galaxies,
whilst another 10\% have colours consistent with BM galaxies.  This mix
of BX/BM is comparable to the observational constraints from C05.

\subsection{Restframe Optical Properties}

\begin{figure*}
\centerline{\psfig{file=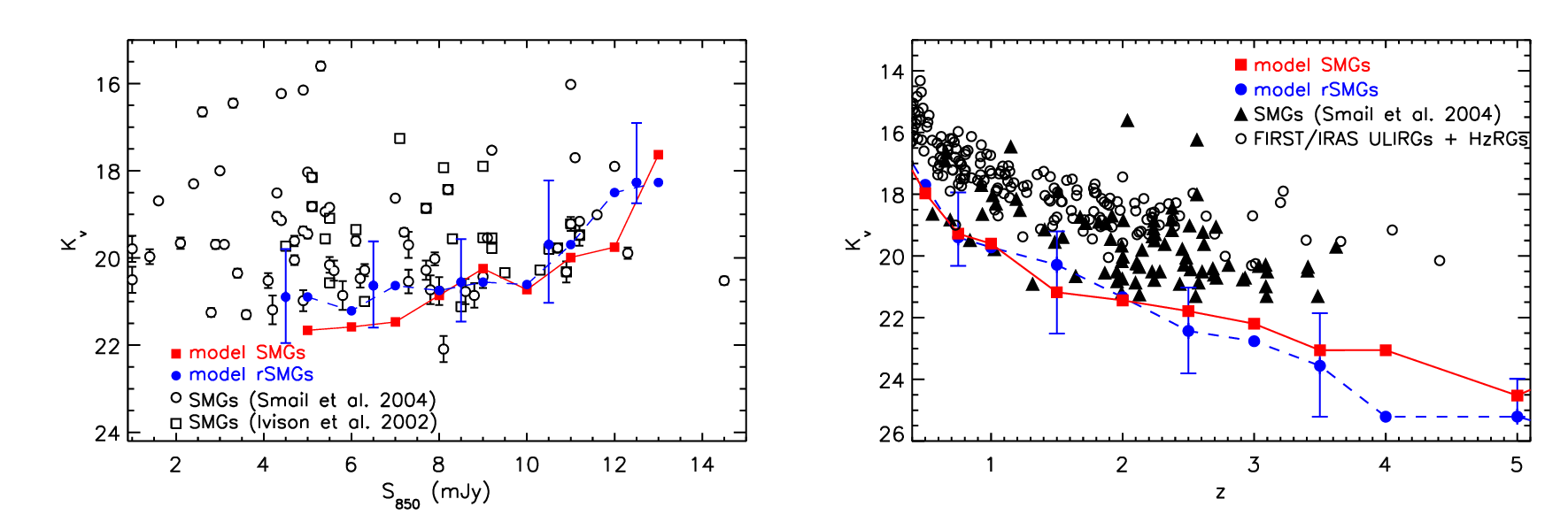,angle=90,width=7.5in}}
\caption{Predicted observed $K$-band magnitude distribution as a
  function of redshift and 850$\mu$m flux density for model SMGs and
  rSMGs compared to the observations (median values are plotted for the
  models).  The error-bars map the central 67\% of the distribution
  from {\sc grasil} and are shown on alternate points for clarity.
  Both plots show that the predicted photometry of model SMGs
  under-estimates that measured from observations by $\sim2$ magnitudes
  ($\sim6\times$).}
\label{fig:Kplots} \end{figure*}

If we turn to longer wavelengths, we can better test the rest-frame
optical properties of galaxies predicted by the models against the
observations.  In Fig.~\ref{fig:Kplots} we compare the predicted
 $K$-band magnitudes of model SMGs with the photometry from
\citet{Smail04}.  Since the stellar population models do not include
emission line components, we have corrected the observed $K$-band
photometry for galaxies between $z=1.8$--2.8 (where the strongest
optical emission line seen in SMGs, H$\alpha$, falls in $K$) for the
H$\alpha$ line emission assuming the median H$\alpha$ equivalent width
of EW$_{rest}$(H$\alpha)=75\pm25$\,\AA\ measured in SMGs
\citep{Swinbank04}. The correction to the $K$-band photometry is
$\lsim10$\%. In contrast to the optical bands, as Fig.~\ref{fig:Kplots}
shows, the $K$-band magnitudes of observed SMGs are brighter at a given
850$\mu$m flux density than all but the brightest model SMGs.  However,
this plot hides the redshift evolution of the $K$-band magnitude and so
a more useful comparison comes from the evolution of $K$-band flux as a
function of redshift. As Fig.~\ref{fig:Kplots} also shows, at $z\sim2$
the median observed $K$-band magnitude for SMGs is $K_{v}=20.0\pm0.3$,
which is approximately two magnitudes brighter than the model
prediction of $K_{v}=22.0\pm0.9$ (for both model SMGs and rSMGs).

\subsection{Restframe Near-infrared Properties}

However, at $z\sim2$, the observed $K$-band samples the rest-frame
$V$-band which is dominated by young stellar populations. So this
comparison is still sensitive to the precise geometry and degree of any
dust obscuration -- where the models may lack sufficient detail. A more
robust test of the predicted photometry from model galaxies comes from
a comparison in the rest-frame $K$-band where the effects of dust
extinction are much reduced. The rest-frame near-infrared can also
provide a better estimate of the mass of the stellar population. At
$z\sim2$, the rest-frame $K$-band is redshifted to $\sim6\mu$m, and
as such the {\it Spitzer Space Telescope} has provided unique insight
into the stellar masses of far-infrared luminous galaxies.

In particular, \citet{Borys05} (see also \citealt{Alexander08})
estimated rest-frame 2.2$\mu$m luminosities for a sample of
spectroscopically confirmed SMGs in the GOODS-N region from broad-band
photometry covering $UBVRIzJK$+3.6/4.5/5.8/8$\mu$m which spans the
rest-frame UV, optical and near-infrared at the redshifts of their
sample, $z=0.6$--2.9. They derive $M_{K}=-26.8\pm0.4$ from their sample
of ten galaxies at $z>1.5$. They then estimate a light-to-mass ratio
for the stellar population of L$_{K}/$\,M\,$=3.2$ using the Starburst99
model \citep{Leitherer99} based on an assumed mean age of 200\,Myr and
a Miller-Scalo IMF \citep{MillerScalo79} and so infer a typical stellar
mass of $\log($M$_{\star})=2.5^{+3.8}_{-2.5}\times 10^{11}\Msol$.
\citet{Borys05} estimate that the dust extinction in the rest-frame
$K$-band is only $\sim 0.2$mag for their galaxies, and so do not
correct for it in their stellar mass estimates.  In comparison, our
model SMGs have median stellar masses
$M_{\star}\sim2.1^{+3.0}_{-1.0}\times10^{10}\Msol$, up to an order of
magnitude lower than inferred by \citet{Borys05} (we note that the
stellar masses quoted here from both observations and the models
include the mass of living stars plus remnants).

\begin{figure*}
\centerline{\psfig{file=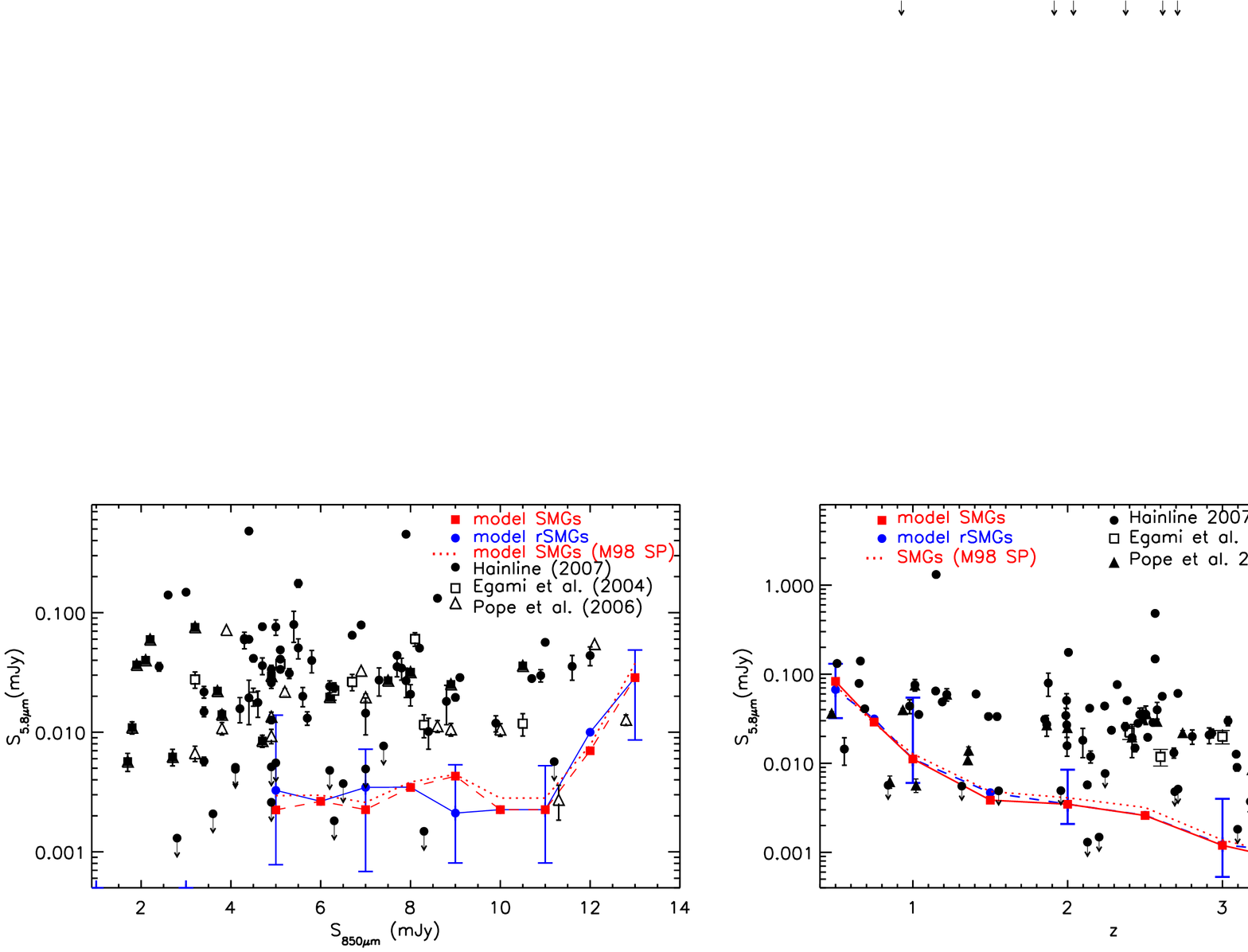,angle=90,width=7.5in}}
\caption{Predictions for the 5.8$\mu$m flux density of model SMGs as a
  function of 850$\mu$m flux density and redshift.  In both panels we
  show the trend of the predicted photometry from {\sc grasil} and the
  changes which occur when the star-formation histories assume the
  stellar population synthesis from \citet{Maraston98} which include a
  full treatment of TP-AGB stars.  The comparison sample comprise SMGs
  with (secure) spectroscopic redshifts from
  \citet{HainlinePhDTh,Egami04} and \citet{Pope06}.  Both plots show
  that the predicted 5.8$\mu$m flux density for model SMGs is
  systematically lower than the observations (by a factor 10$\times$),
  and we suggest that this deficit is due to the lower-than-expected
  stellar mass for model SMGs ($M_{\star}\sim10^{10}\Msol$ whilst
  observations suggest more massive systems,
  $M_{\star}\sim10^{11}\Msol$)}
\label{fig:IRAC58plots}
\end{figure*}

However, since deriving the stellar mass is sensitive to the
mass-to-light ratio (which depends on the assumed IMF, as well as the
age of the stellar population and the dust extinction), a more robust
comparison is to simply compare the observed mid-infrared flux
densities. In Fig.~\ref{fig:IRAC58plots} we compare the predicted
5.8$\mu$m flux density with observations compiled from deep {\it
Spitzer} surveys \citep{Egami04,HainlinePhDTh,Pope06}.  For $z=2$--3,
observations suggests a median 5.8$\mu$m flux density of
$35\pm8\mu$Jy. In comparison, the model SMGs have 5.8$\mu$m flux
density of $3.6^{+8.0}_{-2.4}\mu$Jy, thus under-predicting the
5.8$\mu$m flux density by a factor $\sim10\times$. Thus the low
rest-frame $K$-band fluxes in the models compared to observations seem
to point to the same conclusion about the stellar masses in the model
being too low.  

We note that the rest-frame stellar mass-to-light ratio for the model
SMGs (including dust extinction) is L$_{K}/$\,M\,$=4.8\pm2.9$. This is
not greatly dissimilar to the value used by \citet{Borys05}, but given
the different IMFs, stellar ages and dust extinctions assumed, this
apparent consistency may be somewhat fortuitous.

This effect of the predicted 5.8$\mu$m fluxes being lower than
observed values is not only limited to the model SMGs. We also use the
model to select Lyman-Break Galaxies which have also had extensive
mid-infrared follow-up. The median 5.8$\mu$m flux density of 72
spectroscopically confirmed LBGs between $z=1.5$--3 from
\citet{Shapley05} is $S_{5.8}=2_{-1.5}^{+0.5}\mu$Jy (including non
detections).  In contrast the median 5.8$\mu$m flux density for model
LBGs selected using the same colour cuts and redshift distribution is
$S_{5.8}=0.4^{+0.3}_{-0.2}\mu$Jy, approximately a factor 4$\times$
lower than observations suggest.

Are the low $K$-band and 5.8$\mu$m flux densities for model SMGs an
effect of the stellar population models rather than an effect of the
stellar masses (i.e.\ is there some way that the models can produce
more rest-frame $K$-band light for the same stellar mass)?

One potential route to reconciling the models with observations is to
introduce a full treatment of the thermally-pulsating asymptotic giant
branch (TP-AGB) phase into the stellar populations. Recent stellar
population synthesis modeling by \citet{Maraston06} has shown that
TP-AGB stars in post-starburst galaxies can contribute significantly to
the rest-frame $K$-band luminosity on $\gsim200$--1000\,Myr timescales.
Indeed, the predicted stellar light-to-mass ratio at 250\,Myr for a
single stellar population is L$_K$/M$\sim\,6$--8
(\citealt{Maraston98,Maraston06}). This acts to reduce
the observationally estimated stellar mass by a factor of up to
3$\times$ if the luminosity weighted stellar populations in the SMGs
are dominated by stars with ages of $\gsim200$--1000\,Myr. However,
since sub-mm galaxies are likely to be sustained bursts, the dominant
stellar population is likely much younger, and the mass-to-light ratio
evolves as a function of the burst age.  We note that the median age
of a burst in a sub-mm galaxy in the models is $55_{-30}^{+250}$\,Myr
(where age denotes the time elapsed since the beginning of the burst).
To test the likely effect on the {\sc grasil} luminosities we take the
existing star-formation histories and construct the composite stellar
population using the \citet{Maraston98} stellar population models. We
use a flat IMF ($x=0$) in bursts and standard IMF in the quiescent
mode, as in our fiducial model. The predicted $K$-band and 5.8$\mu$m
flux densities are not significantly effected: the observed $K$-band
flux density increases by $\sim40$\% at $z=2$ whilst the 5.8$\mu$m flux
density increases by only $\sim20$\% on average
(Fig.~\ref{fig:IRAC58plots}).  This suggests a median 5.8$\mu$m flux
density for model SMGs of $\sim4\mu$Jy for the Maraston stellar
population models compared to $3\mu$Jy for the Padova models at
$z=2$--3, still a factor 8--$10\times$ lower than that inferred
observationally.  Thus, although the TP-AGB phase may have a small
effect in the SED modeling, for extended bursts this phase is unlikely
to account for the low $K$-band and 5.8$\mu$m luminosities predicted
for model SMGs.

Another possibility to reconcile the models would be to reduce the
burst lifetimes such that the rest-frame near-infrared light becomes
dominated by TP-AGB stars. For short bursts (i.e.\ $<10$\,Myrs), the
younger luminosity-weighted age for the stellar population in the SMGs
results in a much higher light-to-mass ratio, with
L$_{K}/$\,M\,$=20$--50 at 15--30\,Myr \citep{Maraston06}. However,
these shorter bursts are effectively ruled out since the amount of
energy required to heat the dust in the sub-mm phase remains fixed,
shorter burst durations result in hotter dust temperatures and
therefore the 850$\mu$m sub-mm counts are not matched.  This leads us
to conclude that the deficit of stellar mass is a real effect and can
not be easily explained by either the introduction of the TP-AGB phase
into the stellar population models or by dramatically reducing the
burst lifetime. We return to this in \S5.

One other potential contributor to the low predicted mid-infrared
fluxes is the effect of dust extinction in the model. Model SMGs at
$z\sim2$ typically have extinctions of 4\,mag in the observed $R$-band,
2.4\,mag in the observed $K$-band, and 1.6\,mag at observer-frame
5.8$\mu$m. This means that the unextincted 5.8$\mu$m flux would be
$4\times$ larger than the extincted value.  If the 5.8$\mu$m dust
extinction could be drastically reduced in the model SMGs, while still
retaining a large enough extinction in the rest-frame UV to cause most
of the UV luminosity to be reprocessed by dust, then this could remove
some of the discrepancy between predicted and observed 5.8$\mu$m fluxes
for SMGs, although a significant offset, $\gsim 2$, would remain.
However, we note that there is support for these high extinctions in
the $K$-band, at least for the emission-line gas.  Indeed, studies of
the Balmer decrement in SMGs suggest extinctions of order
$A_v$=2.9$\pm$0.5 \citep{Takata06}.  We do not explore this possibility
further here, but defer this to a future paper.

Finally, we note that while AGN are ubiquitous in SMGs, their typical
contribution to the bolometric emission is $\lsim10$\%
\citep{Alexander05a,Menendez07,Alexander08} suggesting that any
contamination to the rest-frame 2.2$\mu$m flux from AGN activity is
minimal and unlikely to affect the observed stellar masses. Indeed, as
\citet{Borys05} point out, the shape of the spectral energy
distribution is well-described by a stellar model and remove from their
analysis the small number of galaxies for which AGN potentially
dominate the rest-frame near-infrared.

\section{Masses and Evolution of SMGs}

\subsection{Kinematics and Dynamical Masses}

Deriving dynamical masses for high redshift galaxies relies on
measuring rotation curves and/or line widths and sizes for the line
emitting regions. The most reliable line width estimates for sub-mm
galaxies come from resolved dynamics traced through millimetric CO
emission \citep{Greve05,Tacconi06,Tacconi08} and nebular emission (such
as H$\alpha$; \citealt{Swinbank04,Swinbank06b}).  Converting these line
widths into dynamical masses is particularly problematic at high
redshift since the sizes of the galaxies are poorly constrained, even with
$HST$ resolution.  We therefore begin by comparing the line widths of
the model and observed SMGs.  Based on observations of approximately 30
SMGs, $\sigma_{H\alpha}=170\pm30\kms$ and $\sigma_{CO}=200\pm45\kms$
have been derived \citep{Swinbank04,Greve05,Tacconi06,Tacconi08}.  For
the model galaxies the velocity dispersions and sizes are calculated
assuming (i) conservation of angular momentum during the collapse of
the dark matter halo and baryons; (ii) the size of a stellar spheroid
remnant produced by mergers or disk instabilities is determined by
virial equilibrium and energy conservation (see \citealt{Cole2k} for a
detailed discussion). Disks are assumed to have an exponential surface
density profile with half mass radius $r_{disc}$, whilst the spheroid
follows an $r^{1/4}$ law in projection with half mass radius (in 3D)
$r_{bulge}$. The mass distribution in the halo and the length scale of
the disk and bulge are assumed to adjust adiabatically in response to
each other: for the disk the total angular momentum is conserved whilst
for the spheroid $rV_c(r)$ is conserved at $r_{bulge}$. For each galaxy
this results in a size and velocity dispersion for the spheroid and a
size and rotational velocity for the disk and have been tested against
observations by \citet{Cole2k} for local disks and \citet{Almeida07}
for local spheroids.

For both model SMGs and rSMGs we find $\sigma_{1D}=160\pm30\kms$ which
is in good agreement with the observational constraints. In order to
constrain dynamical masses we also need to include the sizes. However,
observationally these are poorly constrained, and usually are taken to
be 4--8\,kpc (which is approximately the size seen in {\it HST}
observations and in resolved spectroscopic imaging). Using simple
dynamical models, observations suggest dynamical masses of order
$\sim2$--$5\times10^{11}\Msol$ within 4--8\,kpc \citep{Swinbank06b}.

\begin{figure}
\centerline{\psfig{file=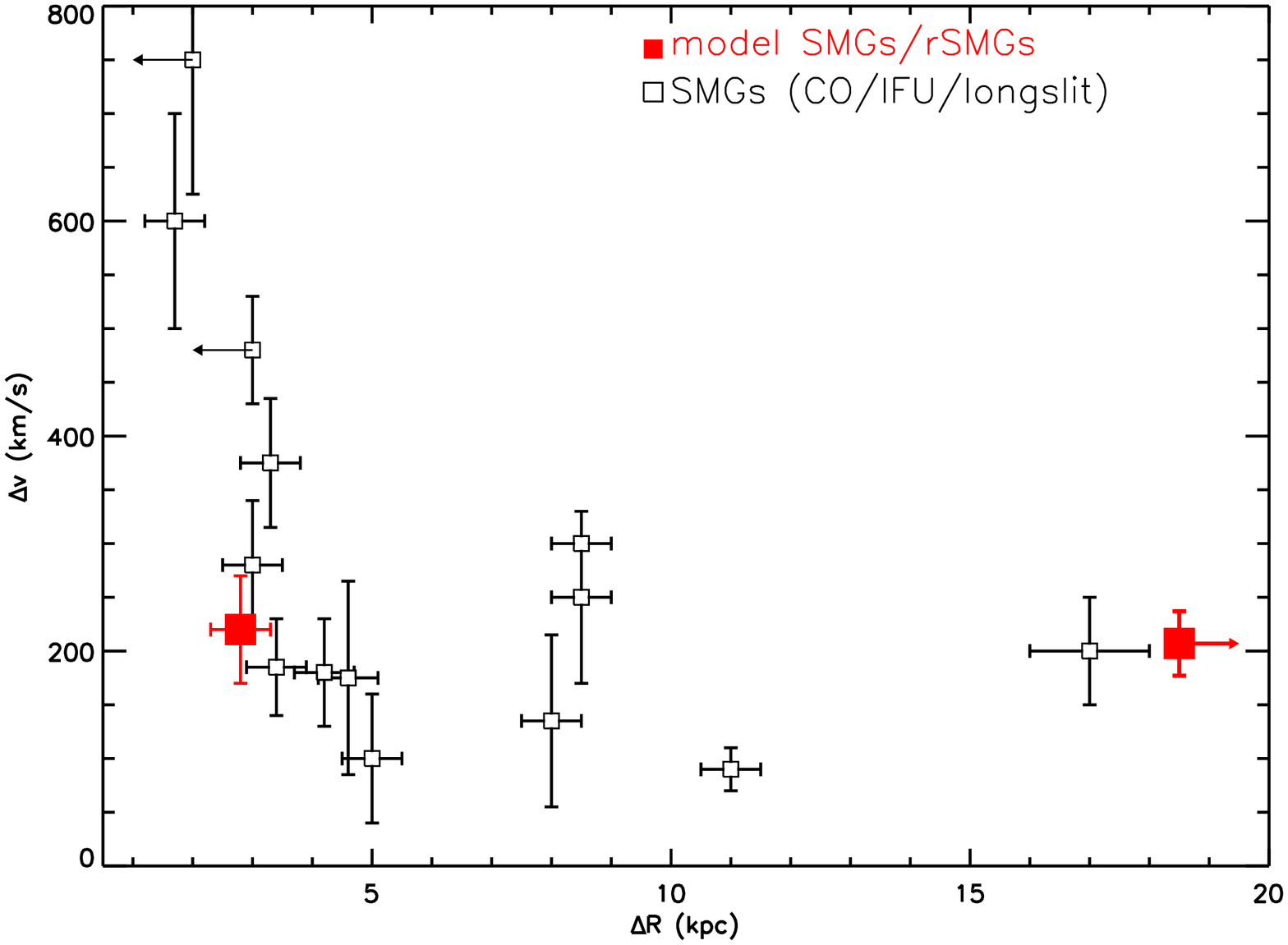,angle=90,width=3.0in}}
\caption{Velocity offset versus spatial offset for the multiple
  components within sub-mm galaxies from IFU, longslit or CO
  observations
  \citep{Tecza04,Greve05,Swinbank04,Swinbank06b,Tacconi06,Tacconi08}.
  We plot the average radius and velocity of the disk for SMGs and
  rSMGs from {\sc galform} showing that the predicted dynamical masses
  of SMGs are in reasonable agreement with those inferred
  observationally.  We also show the median halo circular velocity and
  radius for model SMGs (210$\pm$30$\kms$ and 110$\pm$20\,kpc
  respectively) at the edge of the plot.}
\label{fig:dVdR}
\end{figure}

Since {\sc galform} predicts the size and rotational velocity (or
velocity dispersion) of the bulge and disk, it is also possible to
crudely compare the predicted dynamical masses.  In Fig.~\ref{fig:dVdR}
we show the variation of projected 1-dimensional velocity with spatial
scale for sub-mm galaxies which show multiple components in their
spatially resolved spectra, either from CO, IFU or longslit
observations
\citep{Tecza04,Swinbank04,Swinbank06b,Greve05,Tacconi06,Tacconi08} and
overlay the predicted trend of rotational velocity versus size for
model SMGs.  Although it is clearly difficult to compare the velocity
offsets between merging components and the rotational velocity of
disks, since the predictions for model SMGs lie within the scatter of
the observations, this suggests that the model and observational dynamical
masses are in reasonable agreement (moreover, in an ideal system the
expected difference between the velocity offset between merging
components and the rotational velocity of a disk is a factor
$\sim2\times$).  However, this comparison would benefit from
improvements in both the observations and theoretical predictions (e.g.\
laser guide star adaptive optics integral field spectroscopy would
trace the line emitting regions on sub-kpc regions) coupled with a
theoretical study of the distribution of detectable emission line gas
within dusty, merging systems at high redshift (e.g.\ from a direct
numerical simulation; \citealt{Okamoto08}).

\subsection{Gas Masses}

At its most basic level, the star formation in SMGs is fueled from the
reservoirs of cold H$_2$ gas in these systems. Observations of emission
from CO rotational transitions are the primary means of tracing
(dipole-less) molecular hydrogen, H$_{2}$. The conversion between CO
line luminosity and total cold gas mass is usually expressed via
M$_{gas}$/L$'_{CO}=\alpha$\,M$_{\odot}$\,(K\,km\,s$^{-1}$\,pc$^{2}$)$^{-1}$.
The standard Milky Way ratio is
M(H$_{2}$+He)/L$'_{CO}=4.6$\,M$_{\odot}$\,(K\,km\,s$^{-1}$\,pc$^{2}$)$^{-1}$.
However, in the extreme environment near the center of an
ultra-luminous galaxy, where the CO emission originates from an
inter-cloud medium that is essentially volume filling, rather than from
clouds bound by self-gravity, the CO luminosity traces the molecular
mass as a whole. Extensive high resolution CO mapping of local
($z<0.1$) ULIRGs has been used to trace their gas reservoir sizes and
dynamics to derive an approximate ratio of
M$_{H_{2}+He}$/L$'_{CO}\sim0.8$ \citep{Downes98,SolomonVandenBout05}.
Adopting this value to convert the CO emission line luminosities of
high redshift sub-mm galaxies suggests gas masses
$3.0\pm1.6\times10^{10}\Msol$ within the central 2\,kpc
\citep{Greve05,Tacconi06,Tacconi08}. Note that if $\alpha>1$ and
assuming random orientations then the H$_{2}$ gas mass would exceed the
dynamical mass on average.

In the model SMGs, the median gas mass
M$_{gas}=3.4_{-1.7}^{+2.7}\times10^{10}\Msol$.  A direct comparison of
the gas masses of observed SMGs with those in the models is difficult
due to the uncertainty in the conversion between CO(3--2) and CO(1--0)
line luminosity as well as the conversion between L$'_{CO}$ and
M$_{gas}$ \citep{Hainline06}. Nevertheless, we can at least test what
the value of $\alpha$ would be given the predicted gas masses and
observed line luminosities. The median line luminosity of $>$5\,mJy
SMGs from \citet{Greve05} and \citet{Tacconi06} (including non
detections) is
L$'_{CO}=3\pm1\times10^{10}$\,K\,km\,s$^{-1}$\,pc$^{-2}$. Assuming the
gas mass of model SMGs and rSMGs of
M$_{gas}=3.4_{-1.7}^{+2.7}\times10^{10}\Msol$ and using
$\alpha=$\,M$_{gas}$/L$'_{CO}$ we would require
$\alpha=1.1_{-0.6}^{+1.0}($\,K\,km\,s$^{-1}$\,pc$^{-2}$)$^{-1}$ to
reproduce the observed L$'_{CO}$, which is in good agreement with the
value used to observationally infer gas masses from CO line luminosities
in similarly active systems.

\subsection{Halo Masses}

\begin{figure}
\centerline{\psfig{file=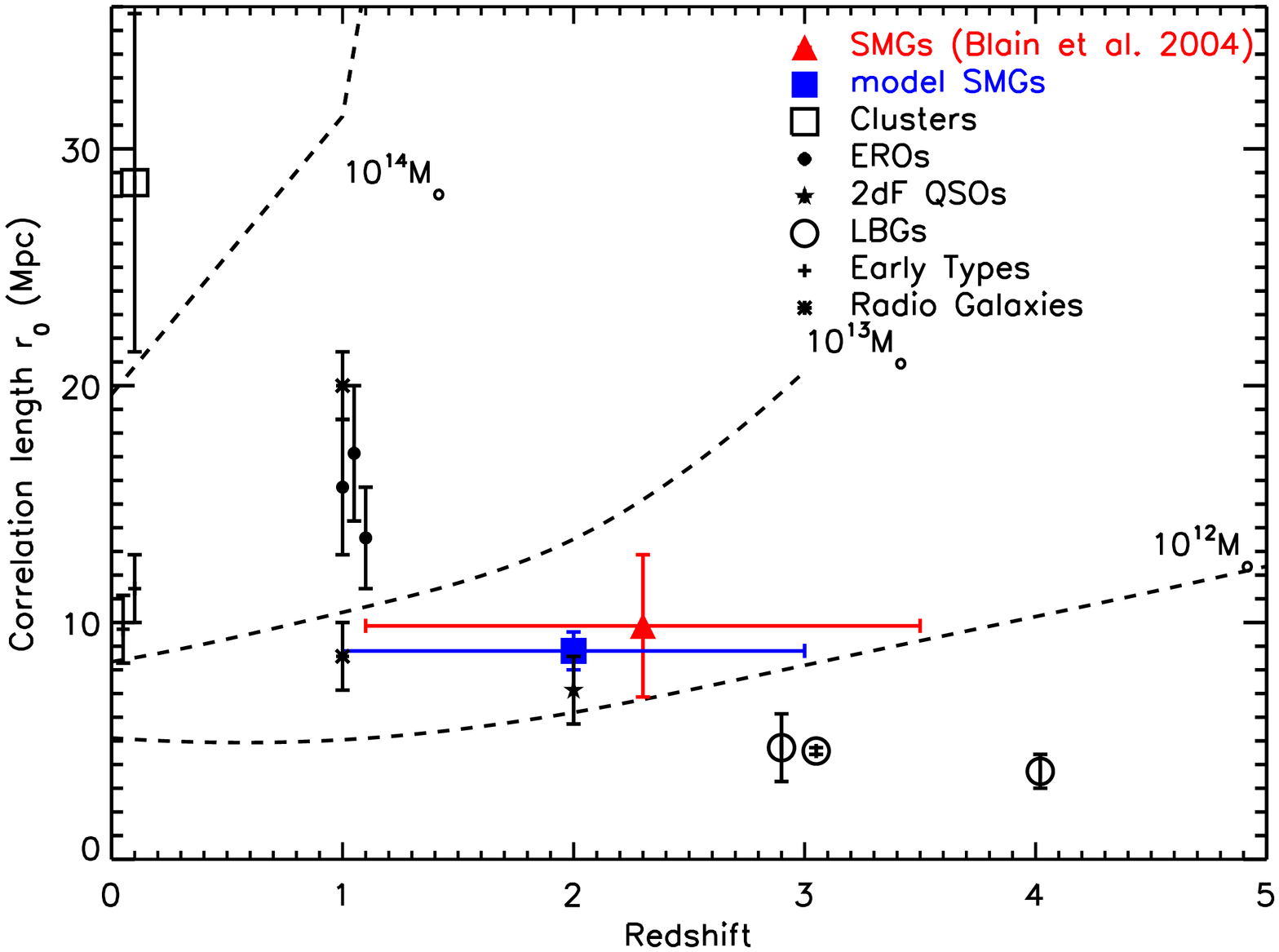,angle=90,width=3.5in}}
\caption{Comoving correlation length for SMGs from \citet{Blain04a}
  compared to model SMGs, as well as in contrast to other populations
  of low- and high-redshift galaxies
  \citep{Bahcall83,McCarthy01,Croom01,Adelberger00,Willmer98,Overzier03}.
  The dashed lines show the expected correlation length of dark matter
  halos as a function of mass and redshift, as measured from the
  Millennium Simulation (Gao et al.\ 2005).  A clustering length of
  $r_0=9.8\pm3.0$\,Mpc corresponds to a halo mass of
  M$_{h}=3.1^{+5.7}_{-1.9}\times10^{12}$\,M$_{\odot}$.  This mass is
  close to that directly predicted for the model SMGs, suggesting that
  any merger bias is low in our model.}
\label{fig:millenium_corr}
\end{figure}

The clustering of sub-mm galaxies encodes important information on the
underlying dark matter distribution (as well as the distribution of
SMGs within and between the halos). A measurement of clustering
therefore provides key constraints on models of galaxy formation.
Although clustering measurements for SMGs are tentative,
\citet{Blain04a} showed the potential of using precise spectroscopic
redshifts to detect clustering in modest-sized samples, deriving a
crude constraint on their clustering length of $r_0=9.8\pm3.0$\,Mpc
(Fig.~\ref{fig:millenium_corr}).  Using the the Millennium Simulation,
Almeida et al.\ (2008 in prep.)  predict that the correlation function
of model SMGs from the same model used here is close to a power law of
the form $\xi=(r/r_{o})^{\gamma}$ over more than three decades in
separation, with $\gamma=-1.94\pm0.05$ and a correlation length
$r_{o}=8.8\pm0.3$\,Mpc, in excellent agreement with observations.

In Fig.~\ref{fig:millenium_corr} we show the clustering lengths of
halos of mass log(M$_{h}$/M$_{\odot})=11$, 12, 13 measured from the
Millennium Simulation for the evolution of the dark matter in the
$\Lambda$CDM model \citep{Springel05}.  For a clustering length of
$9.8\pm3.0$\,Mpc the corresponding halo mass is predicted to be
M$_{h}=3.1^{+5.7}_{-1.9}\times10^{12}$\,M$_{\odot}$ \citep{Gao05}.
This compares to the predicted halo mass for the model SMGs of
M$_{halo}=3.6_{-1.5}^{+5.5}\times10^{12}$\,M$_{\odot}$.  These results
confirm the high masses of the SMG host halos in the model.  Moreover,
the close similarity between the halo masses predicted from their
clustering and the actual halo masses of the model SMGs implies that
any ``merger bias'' has only a weak effect on their clustering
\citep{Percival03}.

\subsection{Comparison to other galaxy populations}
\label{sec:other_pops}

The {\sc galform} model allows us to compare the SMGs and rSMGs with
other high-redshift populations. In this section we briefly discuss the
properties of radio unidentified SMGs, as well as optically faint radio
galaxies (OFRGs) which have been proposed to have a similar redshift
distribution and bolometric luminosities as radio-detected sub-mm
galaxies but with comparatively higher dust temperatures
\citep{Chapman04c,Blain04b} as opposed to cooler temperatures (or
higher redshifts) for the radio undetected SMGs.

To investigate the properties of the model radio undetected SMGs and
their relation to the model SMGs, we select model galaxies with
S$_{850}>5$\,mJy and S$_{1.4}<30\mu$Jy.  These have a median redshift
of $<\! z\!  >=2.4$, $\sigma_z=0.5$, peaking at somewhat higher
redshift than the model SMGs ($<\! z\!  >=2.0$).  The bolometric
luminosities of the model radio undetected SMGs are comparable to the
model SMGs (L$_{bol}=1.6^{+0.8}_{-1.0}\times10^{12}$\,L$_{\odot}$),
with characteristic dust temperatures slightly colder than model SMGs
(T$_{d}=28\pm8$\,K). This supports the suggestion of \citet{Chapman05a}
that the radio undetected SMGs are likely to represent the cold wing of
the distribution of dust temperatures, but with a similar bolometric
luminosity to the radio-detected SMGs.  As noted before, this results
in a modest bias against the highest redshift SMGs in radio-identified
samples.

The halo, stellar and gas masses of this population are
M$_{halo}=2.5^{+7.5}_{-0.8}\times10^{12}$,
M$_{\star}=1.5^{+1.7}_{-0.9}\times10^{10}$ and
M$_{gas}=3.4_{-1.5}^{+1.0}\times10^{10}\Msol$ respectively. Thus the
properties of the model radio undetected SMGs significantly overlap
those of model SMGs, with the main difference arising in their dust
temperatures, where the model radio undetected SMGs have marginally
colder dust temperatures and similar bolometric luminosities, resulting
in a slightly higher redshift distribution compared to model rSMGs.

Similarly, we investigate the properties of the model OFRGs: we define
OFRGs as having $S_{850}<3$\,mJy, $\srad\,>30\mu$Jy and $I_{AB}>22.5$
to be consistent with \citet{Chapman04c}. We find that the model OFRGs
have a much lower mean redshift than the the model SMGs, with $<\! z\!
>=1.2$, $\sigma_z=0.6$.  Since these galaxies are selected via their
radio flux density, their 850$\mu$m/1.4GHz flux density is lower than
the model SMGs with S$_{850}$/S$_{1.4}=17_{-6}^{+5}$ at $z=1$ compared
to model SMGs with S$_{850}$/S$_{1.4}=50^{+50}_{-30}$ at the same
redshift). The halo, stellar and gas masses of OFRGs are
$8^{+16}_{-5}\times10^{11}$, $4_{-3}^{+5}\times10^{9}$ and
$1.5_{-0.5}^{+0.4}\times10^{10}\Msol$ respectively, and bolometric
luminosities of order $6.5\pm1.5\times10^{11}$\,L$_{\odot}$,
approximately 3--5$\times$ lower than the SMGs.

\begin{figure*}
\centerline{\psfig{file=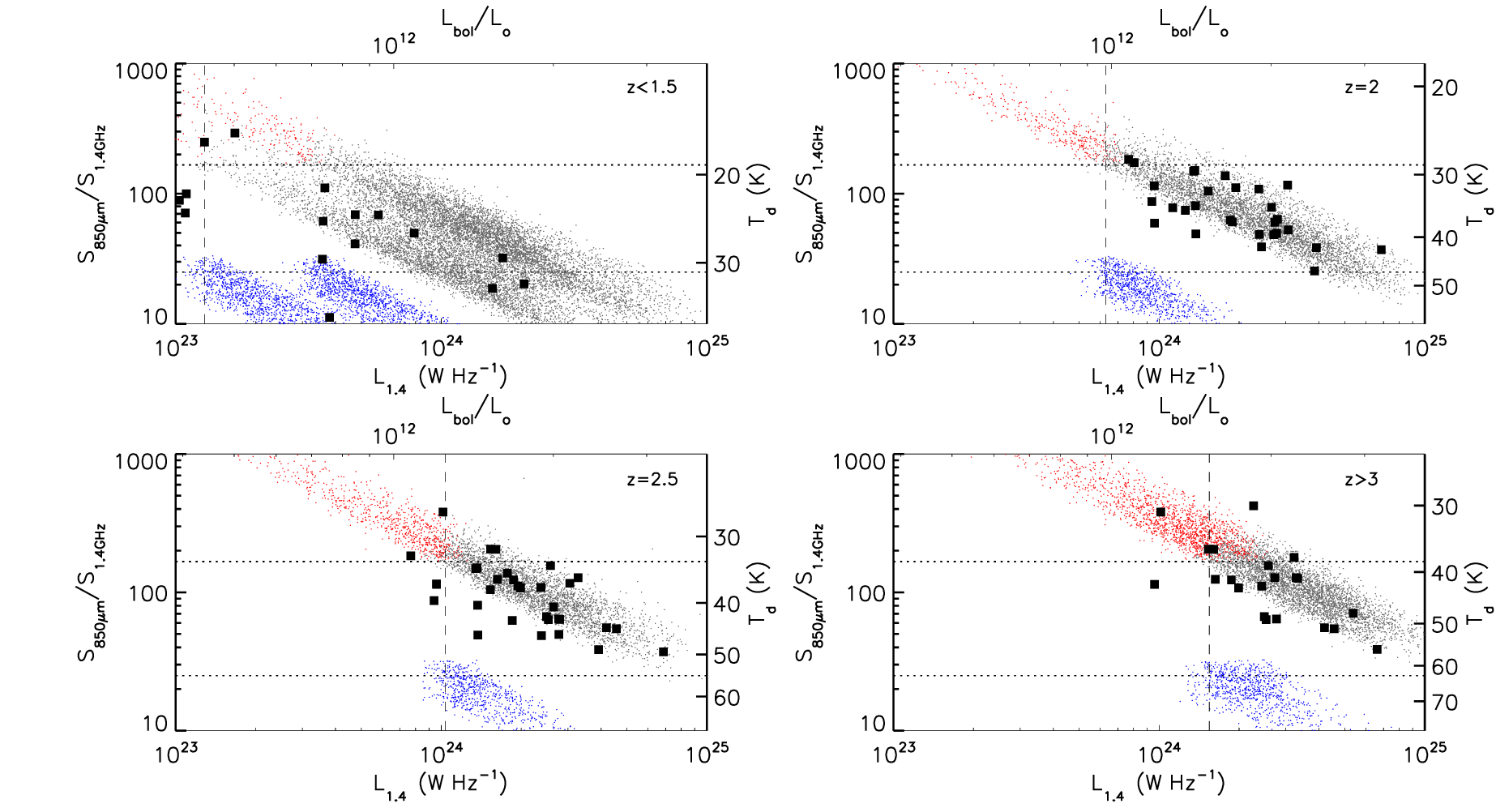,angle=0,width=7.0in}}
\caption{Radio (1.4\,GHz) luminosity versus far-infrared/radio flux
  density ratio as a function of redshift for model SMGs, radio
  undetected SMGs (ruSMGs) and OFRGs compared to SMGs from
  \citet{Chapman05a}.  For clarity we split the data into four
  redshift bins.  We also show the approximate corresponding
  bolometric luminosity, computed using the far-infrared--radio
  correlation defined by the model SMGs.  We also show the approximate
  characteristic temperature.  The plot shows that at a fixed
  luminosity, the model radio undetected SMGs have systematically
  lower temperatures than the model SMGs at each redshift.  We also
  note that (by selection) the model OFRGs have systematically higher
  temperatures at a fixed luminosity compared to the SMGs at all
  redshifts.}
\label{fig:Lbol_Td}
\end{figure*}

To relate the properties of model SMGs, radio undetected SMGs and
OFRGs in Fig~\ref{fig:Lbol_Td} we show the predicted
850$\mu$m/1.4\,GHz flux density ratio as a function of 1.4GHz
(rest-frame) radio luminosity, split into four redshift bins ($z\leq1.5$,
$z=2$, $z=2.5$ and $z\geq3$). For comparison we overlay the SMGs from
C05. We also show along the top axis the approximate bolometric dust
luminosity corresponding to the radio luminosity, computed using the
far--infrared--radio correlation defined by the model SMGs (which has
the form $log(L_{bol}/L_{\odot})=-14.40\pm0.15+1.12\pm0.02L_{1.4}$).

As expected, at $z<1.5$--2 the model radio-undetected SMGs are cooler
with lower bolometric luminosities (lower dust temperatures and lower
bolometric luminosities combine to give the same apparent submm flux
density as model SMGs). In contrast, at $z>2.5$ the radio flux density
limit cuts into the typical dust temperatures of model SMGs, and so
typical model SMGs (which would have been identified at lower redshift)
are no longer selected, leaving just the tail of model SMGs with the
highest dust temperatures.  The observations also show three very low
redshift, cold sources (4\% of the C05 sample) which are missed in the
models.  These might arise due to mis-identifications (e.g.\ 2 radio
sources within the beam or an unrelated radio source and a radio
undetected SMG within the beam).  It is also possible that the model is
missing a component of extended cool ULIRG emission (so called cirrus
galaxies; e.g. \citealt{Efstathiou03}).  We also note that there there
are a few high redshift sources with low $S_{850}/S_{1.4}$ flux density
ratios and high bolometric luminosities, indicating possible
contamination from AGN in some of the higher-redshift sources (or an
additional radio component which is not included in the models).

Fig.~\ref{fig:Lbol_Td} also shows that the model OFRGs are predicted
to have higher dust temperatures than model SMGs, and a range of
bolometric luminosities from $L_{bol}\sim10^{11-12.5}L_{\odot}$.

Finally, we also use the model to test what fraction of high-redshift
ULIRGs are identified as SMGs. Selecting galaxies with bolometric
luminosities comparable to SMGs ($>2\times10^{12}$L$_{\odot}$), we find
that the model SMGs comprise $\sim50$\% and $\sim40$\% of the ULIRG
population at $z=1.5$--2.0 and $z=2.5$--3.5 respectively.

\subsection{Comparison between the z$\sim$4 and z$\sim$2 SMG populations}

The negative K-correction experienced at 850$\mu$m potentially allows
very high redshift (up to $z\sim8$) galaxies to be detected in the
sub-mm \citep{Blain96}. The properties of the first sub-mm galaxies
($z\gsim4$) and those at $z\sim2$ may be very different as the
Universe doubles its age over this period. For this reason we have
compared and contrasted the predicted properties of the most distant
sub-mm galaxies, at $z\gsim 4$, with the more typical systems at
$z\sim 2$.

\begin{figure}
\centerline{\psfig{file=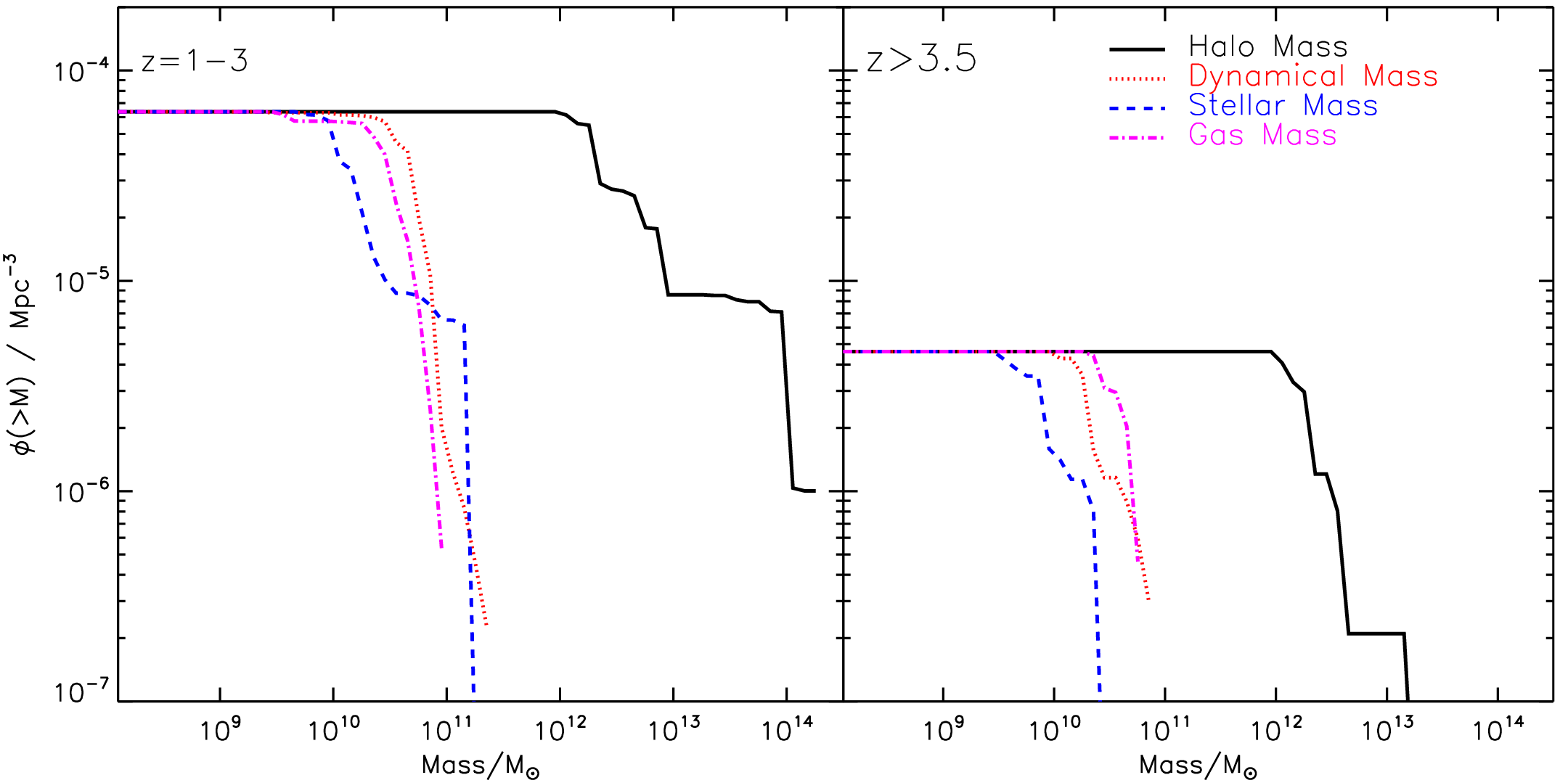,angle=90,width=3.5in}}
\caption{Cumulative mass functions for the halo, dynamical, stellar
  and gas masses for model SMGs split between at $z\sim$2 compared to
  those at $z\sim$4.  The figure shows that the $z\sim2$ model SMGs
  are more ubiquitous by a factor $\sim$6$\times$ compared to those at
  $z\sim4$, but their halo, dynamical, stellar and gas masses are only
  a factor $\sim20$--40\% larger suggesting that the SMG phase occurs
  in similar types of galaxy at both epochs. }
\label{fig:massfn}
\end{figure}

In Fig~\ref{fig:massfn} we split the population into two redshift bins
($z=1$--3 and $z\geq3.5$) and compare the halo, stellar and gas mass
functions. The model SMGs at $z\geq 3.5$ have typical halo, stellar and
gas masses $2.5\times10^{12}$\,M$_{\odot}$,
$1.0\times10^{10}$\,M$_{\odot}$ and 5$\times$10$^{10}$M$_{\odot}$
respectively.  Thus the the $z=1$--3 counterparts have halo and stellar
masses which are a factor 1.25 and 2$\times$ larger respectively, with
the gas reservoirs a factor $\sim$2$\times$ smaller.  The bolometric
luminosities of the $z=1$--3 and $z\geq 3.5$ model SMGs are
L$_{bol}=1.8^{+1.5}_{-1.0}\times10^{12}L_{\odot}$ and
$2.8^{+2.2}_{-1.3}$ $\times$10$^{12}$L$_{\odot}$ respectively. Thus the
lower redshift model SMGs appear to have slightly lower
($\sim1.5\times$) bolometric luminosities, although there is
significant overlap.  This luminosity evolution supports that claimed
trend first noted by \cite{Ivison02} using deep radio and sub-mm
imaging.  We also note that at $z=4$ the radio-detected fraction of
SMGs is $\sim$30\%, which is much lower than at $z=2$ where the radio
detected fraction of SMGs is $\sim95$\%.

Overall, these facts suggest that the $z>3.5$ model SMGs are physically
quite similar to model SMGs at $z=1$--3.  Since there is significant
overlap in their properties, the rapid evolution in the redshift
distribution and space densities of model SMGs between $z\sim2$ and
$z\sim4$ seems at first sight puzzling: the negative K-correction means
that far-infrared luminous galaxies do not get fainter at 850$\mu$m as
their redshift increases. The explanation of their rapid evolution lies
in the hierarchical buildup of structure in the dark matter.  The halo
masses of model SMGs are similar at both epochs
($>10^{12}$M$_{\odot}$), yet the co-moving space density of
$>10^{12}\Msol$ halos at $z\sim4$ is five times less than at $z\sim2$.
This evolution in the halo number density accounts for most of the
change in the number density of model SMGs over this redshift range,
although other factors also enter, such as the fraction of halos
undergoing major mergers \citep{Lacey93}, and the fraction of baryons
contained in the form of cool gas.

\section{Discussion}

Our fiducial model has previously been shown to fit the properties of
local IRAS galaxies, as well as the number counts of faint sub-mm
galaxies.  Here, we have compared the SEDs, dynamical and halo masses
of sub-mm galaxies against observational data, showing good agreement
with the values measured or inferred from observations.  However, the
largest discrepancy between the observations and the predictions from
the model arises in the observed $K$-band and mid-infrared photometry.
Although we find that in the model SMGs the recent starburst dominates
the $K$-band and mid-infrared flux densities, contributing $\sim70$\%
at 5.8$\mu$m and $\sim80$\% in the $K$-band, we find that the $K$-band
fluxes of model SMGs are under-predicted by a factor $\sim\,6\times$
compared to observations, whilst the 5.8$\mu$m fluxes (which
corresponds to rest-frame $K$-band at $z\sim2$) are under-predicted by
a factor $\sim\,10\times$.  This discrepancy is not limited to the
model SMGs, but also applies to other model populations.  For example,
selecting Lyman-Break Galaxies (using $ugr$ colours to select
ubiquitous star-forming galaxies at $z\sim3$) we find that the model
LBGs at $z=2$--3.5 have 5.8$\mu$m flux densities a factor $\sim4\times$
lower than inferred from observations.  We show that including a full
treatment of the TP-AGB phase in the stellar population modeling
\citep{Maraston98} cannot make up for this deficit for any plausible
star-formation history, since this increases the 5.8$\mu$m flux
densities by only $\sim 20\%$ on average. We also note that within the
halo hosting the model SMG, 80--90\% of the stellar mass is associated
with the model SMG itself, thus including more of the stellar mass from
satellites would have little effect on the integrated light.

The low $K$-band and mid-infrared flux densities are a concern for this
model, since this is a key indication of the stellar mass and dust
reddening.  The obvious interpretation is that the stellar masses of
the model SMGs (M$_{star}\sim2\times10^{10}\Msol$) are too low.
However, in the model SMGs the total stellar mass formed in the current
burst is $\lsim10$\% of the total stellar mass at the epoch of
observation (typically 1--$4\times10^{9}\Msol$ built in a period of
$55_{-30}^{+250}$\,Myr which is the median age of the burst in a model
SMG).  Furthermore, the total stellar mass formed in all of the
previous bursts contributes less than 10\% of the total stellar mass of
the model SMG at the epoch of observation.  This means that the burst
mode is responsible for forming less than $\lsim20$\% of the total
stellar mass in model SMGs.  Since the current burst dominates the
luminosity, the rest-frame $K$-band light-to-mass ratio for the
existing- and burst- stellar populations in the model SMGs are very
different, with $L_K/M$=1.5$_{-0.6}^{+5.0}$ and $L_K/M$=14.5$\pm$9.5
for the existing- and burst- stellar population respectively. Since the
current burst is predicted to dominate the rest-frame $K$-band
luminosity, even though it contains only $\sim 10\%$ of the stellar
mass one solutions would be to boost the mass in the current burst by a
factor $\sim10$.  However, if everything else in the model was left
unchanged, this would result in an increase in bolometric luminosity by
the same factor.  However, since the the bolometric dust luminosities
of SMGs already agree with observations in the current model, this will
violate the 850$\mu$m counts.

Our current model assumes a very flat IMF in bursts, with slope $x=0$,
so one could consider changing this slope.  Would steepening the IMF
slope to something closer to the Salpeter value $x=1.35$ (so producing
relatively more lower-mass stars), while at the same time increasing
the total burst mass, allow the model to increase the $K$-band
luminosity while keeping the bolometric dust luminosity equal to
observed values?  To test this we examine the ratio of the
$K$-band-to-bolometric luminosities for both a top-heavy and Salpeter
IMF in bursts, but find that this only weakly depends on the IMF slope
from $x=0$ to $x=1.5$, presumably because both luminosities are
dominated by massive stars at these young ages.  Thus changing the IMF
slope does not seem enough to boost the $K$-band luminosity by a factor
10$\times$ for a fixed bolometric luminosity.

One other potential avenue to increase the predicted near-infrared
fluxes in the model SMGs is to alter the dust extinction.  As expected,
the model SMGs have high dust extinctions, even in the rest-frame
$K$-band, where the extinction is typically about 1.6\,mag.  Therefore
if extinction law could be altered in the model to greatly lower the
extinction in the rest-frame $K$-band, while keeping the extinction
high in the rest-frame UV, then the rest-frame $K$-band luminosities
could be boosted by a factor $\sim 4\times$ while keeping the dust
bolometric luminosities the same. This would go some way towards
resolving the problem with the predicted 5.8$\mu$m flux densities being
too low.  However, we note that there is support for these high
extinctions in the $K$-band, at least for the emission-line gas.
Indeed, studies of the Balmer decrement in SMGs suggest
$A_v$=2.9$\pm$0.5, thus removing all of the dust extinction in the
observed $K$-band may violate observations constraints.  We do not
explore this possibility further here, but defer this to a future
paper.

Alternatively, is it possible to increase the stellar mass by
condensing more gas into stars both prior to and during the sub-mm
phase thus boosting the observed $K$-band and 5.8$\mu$m luminosities?
In order to prevent the over-cooling of baryons into stars (and
therefore match the present day $K$-band luminosity function) the
feedback prescription in the B05 model is particularly efficient, with
an instantaneous feedback rate, $\beta=\dot{M}_{ej}$/SFR\,$=1.1$--2.0
(between $z=1$--3) where $\dot{M}_{ej}$ is the gas ejection rate.  When
recycling from dying stars is included this factor is even more
extreme, especially for a top-heavy IMF. In this case, the the recycled
fraction is $R=0.9$, thus only 10\% of the newly formed stellar mass
ends up in long lived stars and remnants (thus the ratio of mass
ejection rate to formation rate of long lived stars is 10--$20\times$
during an SMG burst). This efficient feedback results in 55--65\% of
the baryonic mass in the hot phase for a model SMG at the epoch of
observation, with a hot gas mass of
M$_{hot}\sim8\times10^{10}$M$_{\odot}$ and a total baryonic mass of
M$_{bary}\sim1.5\times10^{11}M_{\odot}$. Thus if the feedback could be
reduced, and a substantial fraction of the mass contained within both
the cold and hot gas reservoirs could be condensed into stars either
prior or during the burst, the result would be an increase in stellar
mass of a factor $\sim6\times$.  This has the obvious draw-back that
there would be no gas available for future star-formation (unless the
halo accretes gas from a merger).

The most promising route to explaining both the near- and far-infrared
luminosities in sub-mm galaxies may lie in combining prescriptions in
the B05 model with those in \citet{Bower06} which uses AGN feedback to
quench gas cooling in high-mass halos and thus prevent the formation of
too many very high-mass galaxies by the present day, rather than
relying on very strong supernova feedback, as in the B05 model. In the
\citet{Bower06} model, high-mass galaxies form earlier than in the B05
model, so the stellar mass function at high redshift has more high-mass
galaxies (though both have very similar stellar mass functions at
$z=0$), and this gives better agreement with the observed evolution of
the rest-frame $K$-band luminosity function than for B05.  However, the
\citet{Bower06} model in its standard form with a single solar
neighborhood IMF under-predicts the sub-mm counts by a factor
$\sim20\times$.  We have investigated modifying the \citet{Bower06} by
introducing a top-heavy IMF in bursts, as in the B05 model.  We find
that we can obtain a comparably good fit to the 850$\mu$m counts as the
B05 model, but the same model then predicts a present-day $K$-band
luminosity function which is too bright by about 1 magnitude at the
bright end.  However, we have not explored the full parameter space of
models combining AGN feedback with a top-heavy IMF, and so it may be
that a model can be found which explains both the number counts and
rest-frame $K$-band luminosities of the SMGs, while also being
consistent with the observed $K$-band luminosity function at different
redshifts. We will investigate this in more detail in a future paper.

\section{Conclusions}

In this paper we have used the {\sc galform} semi-analytic model for
galaxy formation together with the {\sc grasil} spectrophotometric code
to investigate in detail the properties of sub-mm galaxies in the model
of \citet{Baugh05}. We mimic the observational selection in our model
catalogue and identify galaxies both via their 850$\mu$m and 1.4\,GHz
(radio) emission. We find that the radio-identified SMGs (rSMGs) are
predicted to make up approximately 75\% of the whole population. This
is in good agreement with the fraction found in observational surveys
(eg.  \citealt{Ivison05,Ivison07}). The main difference between the
radio-identified and radio-unidentified sub-mm galaxies is the
characteristic dust temperature, where the radio undetected SMGs are
$\sim10$\% cooler.  This results in model SMGs having a median redshift
of $z=2.0$ whilst model rSMGs (which include a radio flux density
threshold S$_{1.4}>30\mu$Jy) have a median redshift of $z=1.7$.

We show that the far-infrared colours of model rSMGs are consistent
with observational data. Fitting the 350 and 850$\mu$m photometry we
show that the model far-infrared SEDs are well described by single
component modified ($\beta=1.5$) blackbodies with temperature of
$32\pm5$\,K (although {\sc grasil} does not assume a single dust
temperature either within or between galaxies). Integrating the model
SED and the black-body fit, the inferred and true bolometric
luminosities for the model SEDs also agree, with the single modified
blackbody fit recovering an average of 0.94$\pm$0.25 of the bolometric
luminosity.  However, the median 850$\mu$m/1.4\,GHz flux density ratio
for SMGs predicted by the model is systematically lower than that
observed by a factor $\sim1.26\pm0.24\times$, even after the model has
been normalized to match the far-infrared--radio correlation for ULIRGs
at $z=0$.  This could be due to evolution in the far-infrared--radio
correlation, or due to modest contributions to the observed radio flux
density from AGN.

We show that the predicted velocity dispersions are in good agreements
with observations, with $\sigma_{1D}\sim160\kms$ for the model SMGs,
compared to observational constraints of $\sigma=170$--$200\kms$
inferred from the kinematics of the redshifted H$\alpha$ and CO
emission.  We also show that the predicted gas masses are agree well
with those derived from observations using $\alpha=0.8$.  Turning to
the halo masses, we find that the model SMGs reside in halos with
masses of M$_{halo}\sim10^{12.4}$M$_{\odot}$, and a correlation length
of $r_{o}=8.8\pm0.3$\,Mpc, in excellent agreement with the tentative
observational estimate of $r_{o}=9.8\pm3.0$\,Mpc from \citet{Blain04a}.
We find no evidence for a significant merger bias in the clustering of
SMGs in the model.

However, we find that the predicted $K$-band and mid-infrared
($5.8\mu$m) flux densities of the model SMGs are up to a factor
$10\times$ fainter than observed. We discuss the possible reasons for
this. The stellar masses of the model SMGs,
M$_{\star}\sim10^{10}$\,M$_{\odot}$, are a factor 10$\times$ smaller
compared to observational estimates based on the same mid-infrared data
(M$_{\star}\gsim10^{11}$\,M$_{\odot}$), but the observational estimates
are very sensitive to the assumed IMF, burst age and dust extinction.
The simplest explanation for this discrepancy in the near- and
mid-infrared fluxes is that it is due to the stellar masses being too
low in the model. However, other factors may be contributing, for
example in the model SMGs, the dust extinction is quite large even at
$5.8\mu$m, resulting in a $4\times$ reduction in the observed flux. If
the rest-frame visible and near-infrared dust extinctions in the model
SMGs were removed, this would improve (but not eliminate) this
discrepancy (an offset of a factor 2$\times$ would remain).  However,
we note that there is support for these high extinctions in the
$K$-band, at least for the emission-line gas.  We also discuss possible
routes for increasing the near-infrared luminosities and stellar
masses, both within the context of the current \citep{Baugh05} model,
which assumes a top-heavy IMF in bursts, and using the prescriptions
described in \citet{Bower06}, which uses AGN feedback to prevent too
many baryons cooling in massive halos but assumes a normal IMF
throughout. The \citet{Bower06} model predicts more high-mass galaxies
at high redshift, and a $K$-band luminosity function in better
agreement with observations at high redshift, but under-predicts the
sub-mm counts by a factor 20. We will discuss the properties of sub-mm
galaxies in models which combine features from the \citet{Baugh05} and
\citet{Bower06} models in a future paper.

We also investigate the evolution of the sub-mm galaxy population in
the \citet{Baugh05} model by comparing the halo and stellar masses for
model SMGs at $z=1$--3 to those at $z>3.5$.  The $z>3.5$ SMG
population are predicted to make up approximately 2\% of the total SMG
population (with the $z>4$ rSMGs contributing approximately 0.5\% of
the total population).  Nevertheless, we find only a mild increase in
the halo and stellar masses between $z\sim4$ and $z\sim2$, and
therefore attribute the strong evolution seen in the redshift
distribution primarily to the evolution in the space density of
massive halos hosting SMGs.

As well as the need for the models to provide a better match to
observations, a number of upcoming surveys and facilities will provide
valuable observational constraints which can be used to further test
them.  In particular, deep SCUBA2 surveys will probe both 450 and
850$\mu$m, providing much tighter constraints on the SEDs of galaxies
with bolometric luminosities a factor $10\times$ lower than currently
achievable. When combined with results from {\it Herschel} (which will
probe 60--670$\mu$m), the temperatures, bolometric luminosities and
dust masses of SMGs will be constrained with much higher accuracy than
currently possible from the sparse and low signal-to-noise 350 and
850$\mu$m photometry alone.  Interestingly, the model predicts that
there is a significant tail of faint 450$\mu$m selected galaxies at
high redshift (Lacey et al.\ 2009 in prep).  Selecting galaxies with
S$_{450}>2.5$\,mJy (the expected 5-$\sigma$ flux density limit for
SCUBA2) the model suggests that there should be $\geq8000$, 850 and
150 galaxies selected at 450$\mu$m per square degree at $z>2$, 4 and 5
respectively ($\sim50$\%, 5\% and 1\% of the total population
respectively), suggesting that the fainter confusion limit at
450$\mu$m may provide the most efficient route to get an unbiased
census for the far-infrared selected galaxies at the highest
redshifts, providing insights into the physical processes of
star-formation occurring at the peak epoch of galaxy formation,
$z\sim2$.

\section*{Acknowledgments}

We would like to thank the anonymous referee for providing a number of
suggestions which significantly improved the content and clarity of
this paper.  We thank Claudia Maraston for providing us with stellar
population models with a top-heavy IMF, and John Helly and Liang Gao
for calculating the merger rates and clustering lengths of halos from
the Millennium Simulation.  We also gratefully acknowledge Dave
Alexander, Richard Bower, Reinhard Genzel and Linda Tacconi for
valuable discussions.  AMS acknowledges an STFC fellowship.  IRS \& CMB
acknowledge the Royal Society.  This work was also supported by the
STFC rolling grant for extragalactic astronomy and cosmology at Durham.

\bibliographystyle{apj}
\bibliography{/Users/ams/Projects/ref}
\bsp

\end{document}